\documentstyle[aps,twocolumn]{revtex}
\input{epsf}

\title{Encoding a qubit in an oscillator\thanks{CALT-68-2273}}
\author{Daniel Gottesman,$^{(1,2)}$\thanks{\tt gottesma@eecs.berkeley.edu} Alexei
Kitaev,$^{(1)}$\thanks{\tt kitaev@microsoft.com}  and John Preskill$^{(3)}$\thanks{\tt
preskill@theory.caltech.edu}}

\address{$^{(1)}$Microsoft Corporation, One Microsoft Way, Redmond, WA 98052, USA\\
$^{(2)}$Computer Science Division, EECS, University of California, Berkeley, CA 94720, USA\\ 
$^{(3)}$Institute for Quantum Information, California Institute of Technology, 
Pasadena, CA 91125, USA }

\begin{document}

\maketitle

\begin{abstract}
Quantum error-correcting codes are constructed that embed a finite-dimensional code space in the infinite-dimensional Hilbert space of a system described by continuous quantum variables. These codes exploit the noncommutative geometry of phase space to protect against errors that shift the values of the canonical variables $q$ and $p$. In the setting of quantum optics, fault-tolerant universal quantum computation can be executed on the protected code subspace using linear optical operations, squeezing, homodyne detection, and photon counting; however, nonlinear mode coupling is required for the preparation of the encoded states.  Finite-dimensional versions of these codes can be constructed that protect encoded quantum information against shifts in the amplitude or phase of a $d$-state system. Continuous-variable codes can be invoked to establish lower bounds on the quantum capacity of Gaussian quantum channels. 
\end{abstract}

\section{Introduction}
Classical information can be carried by either a discrete (digital) signal or a continuous (analog) signal. Although in principle an analog signal can be processed, digital computing is far more robust -- a digital signal can be readily re-standardized and protected from damage caused by the gradual accumulation of small errors.

Quantum information can also be carried by either a discrete (finite-dimensional) system like a two-level atom or an electron spin, or by a continuous (infinite-dimensional) system like a harmonic oscillator or a rotor.  Even in the finite-dimensional case, quantum information is in a certain sense continuous -- a state is a vector in a
Hilbert space that can point in any direction.  Nevertheless, we have known for nearly five years that cleverly encoded quantum states can be re-standardized and protected from the gradual accumulation of small errors, or from the 
destructive effects of decoherence due to uncontrolled interactions with the environment \cite{shor_9,steane_7}.

One is tempted to wonder whether we can go still further and protect the quantum state of a system described by
{\em continuous quantum variables}. Probably this is too much to hope for, since even the problem of protecting analog classical information seems to pose insuperable difficulties.

In this paper we achieve a more modest goal: we describe quantum error-correcting codes that protect a state of a {\em finite-dimensional} quantum system (or ``qudit'') that is encoded in an infinite-dimensional system. These codes may be useful for implementing quantum computation and quantum communication protocols that use harmonic oscillators or rotors that are experimentally accessible. 

We also explain how encoded quantum states can be processed fault tolerantly. Once encoded states have been prepared, a universal set of fault-tolerant quantum gates can be implemented using, in the language of quantum optics, linear optical operations, squeezing, homodyne detection, and photon counting. However, for preparation of the encoded states, nonlinear couplings must be invoked.

Our continuous-variable quantum error-correcting codes are effective in protecting against sufficiently weak diffusive phenomena that cause the position and momentum of an oscillator to drift, or against losses that cause the amplitude of an oscillator to decay. By concatenating with conventional finite-dimensional quantum codes, we can also provide protection against errors that heavily damage a (sufficiently small) subset of all the oscillators in a code block.

A different scheme for realizing robust and efficient quantum computation based on linear optics has been recently proposed by Knill, Laflamme, and Milburn \cite{lanl_linear,lanl_thresh}.

We begin in \S\ref{sec:shift} by describing codes that embed an $n$-state quantum system in a larger $d$-state system, and that protect the encoded quantum information against shifts in the amplitude or phase of the $d$-state system. A realization of this coding scheme based on a charged particle in a magnetic field is discussed in \S\ref{sec:landau}. Our continuous-variable codes are obtained in \S\ref{sec:sing_osc} by considering  a $d\to\infty$ limit. Formally, the code states of the continuous-variable codes are nonnormalizable states, infinitely squeezed in both position and momentum; in \S\ref{sec:finite} we describe the consequences of using more realistic approximate code states that are finitely squeezed. In \S\ref{sec:many_osc} we outline the theory of more general continuous-variable codes based on lattice sphere packings in higher-dimensional phase space. 

We discuss in \S\ref{sec:error} how continuous-variable codes protect against quantum diffusion, amplitude damping, and unitary errors. In \S\ref{sec:gaussian}, we establish a lower bound on the quantum capacity of the Gaussian quantum channel.

We then proceed to develop schemes for fault-tolerant manipulation of encoded quantum information, starting in \S\ref{sec:symplectic} with a discussion of the symplectic operations that can be implemented with linear optics and squeezing. In \S\ref{sec:recovery} we discuss the measurement of the error syndrome and error recovery, which can be achieved with symplectic operations and homodyne detection. Completion of the fault-tolerant universal gate set by means of photon counting is described in \S\ref{sec:universal}, and the preparation of encoded states is explained in  \S\ref{sec:encoding}.

Finally, \S\ref{sec:physical} contains some further remarks about the physical realization of our coding schemes, and \S\ref{sec:conclude} contains concluding comments.

\section{Shift-resistant quantum codes}
\label{sec:shift}
The main novelty of our new codes is that they are designed to protect against
a different type of error than has been considered in previous discussions of
quantum coding.  This distinction is more easily explained if we first consider not the case of a continuous quantum variable, but instead the (also
interesting) case of a ``qudit,'' a $d$-dimensional quantum system.  Quantum
codes can be constructed that encode $k$ protected qudits in a block of $N$
qudits, so that the encoded qudits can be perfectly recovered if up to $t$
qudits are damaged, irrespective of the nature of the damage \cite{knill_higher,chau,rains,higher}.  Error recovery will be effective if errors that act on many qudits at once are rare. More precisely, a general error superoperator acting on $N$ qudits can be expanded in terms of a basis of operators, each of definite ``weight'' (the number of qudits on which the operator acts nontrivially). Encoded information is well protected if the error superoperator has nearly all its support on operators of weight $t$ or less. 

But consider instead a different situation, in which the amplitude for an error
to occur on each qudit is not small, but the errors are of a restricted type.  The possible
errors acting on a single qudit can be expanded in terms of a unitary operator basis with
$d^2$ elements, the ``Pauli operators'':
\begin{equation}
X^a Z^b, \quad  a,b = 0,1,2,\dots, d -1~.
\end{equation}
Here $X$ and $Z$ are generalizations of the Pauli matrices $\sigma_x$ and $\sigma_z$, which act in a particular basis
$\{|j\rangle, j = 0,1,2,\dots, d-1\}$ according to  
\begin{eqnarray}
X&:& |j\rangle\rightarrow |j + 1~({\rm mod}~d)\rangle~,\nonumber \\
Z&:& |j\rangle\rightarrow \omega^j |j\rangle~,
\end{eqnarray}
where $\omega = \exp (2\pi i/d)$. Note that it follows that
\begin{equation}
ZX = \omega XZ.
\end{equation}
For $N$ qudits, there is a unitary operator basis with $d^{2N}$ elements consisting of all tensor products of single-qudit Pauli operators.

We will now imagine that errors with $|a|, |b|$ small compared to $d$ are common, but
errors with large $|a|$ and $|b|$ are rare. 
This type of error model could be expected to apply in the case of a
continuous quantum variable, which is formally the $d \rightarrow \infty$ limit of a qudit.
For example, decoherence causes the position $q$ and momentum $p$ of a  particle
to diffuse with some  nonzero diffusion constant.  In any finite time interval
$q$ and $p$ will drift by some amount that may be small, but is certainly not zero.
How can we protect encoded quantum information under these conditions?

Fortunately, the general ``stabilizer'' framework \cite{att,gott_stab} for constructing quantum
codes can be adapted to this setting.  In this framework, one divides the elements of a unitary operator basis into two disjoint and exhaustive classes: the set ${\cal E}$ of ``likely errors'' that we want to protect against, and the rest, the ``unlikely errors.''  A code
subspace is constructed as the simultaneous eigenspace of a set of commuting
``stabilizer generators,'' that generate an Abelian group, the ``code stabilizer.''   The code can reverse errors in the set
${\cal E}$ if, for each pair of errors $E_a$ and $E_b$, either $E_a^\dagger E_b$ lies in the stabilizer group, or $E_a^\dagger E_b$ fails to commute with some element of the stabilizer. (In the latter case, the two errors alter the eigenvalues of the generators in distinguishable ways; in the former case they do not, but we can successfully recover from an error of type $a$ by applying either $E_a^\dagger$ or $E_b^\dagger$.) In typical discussions of quantum coding, ${\cal E}$ is
assumed to be the set of all tensor products of Pauli operators with weight up
to $t$ (those that act trivially on all but at most $t$ qudits). But the same principles can be invoked to design codes that
protect against errors in a set ${\cal E}$ with other properties.

Quantum codes for continuous variables have been described previously by
Braunstein \cite{braunstein} and by Lloyd and Slotine \cite{lloyd}.  For example, one code they constructed can be regarded as the continuous limit of a qudit code of the type originally introduced by Shor in the binary ($d=2$) case, an $[[N=9, k=1, 2t +1 = 3]]$ code that protects a
single qudit encoded in a block of 9 from arbitrary damage inflicted on any one
of the 9.  The 8 stabilizer generators of the code can be expressed as
\begin{eqnarray}
&Z_1 Z_2^{-1}~, Z_2 Z_3^{-1}~, Z_4 Z_5^{-1}~, Z_5 Z_6^{-1}~, Z_7 Z_8^{-1}~,Z_8
Z_9^{-1}~,\nonumber\\
&(X_1 X_2 X_3)\cdot (X_4 X_5 X_6)^{-1}~, (X_4X_5X_6)\cdot (X_7X_8X_9)^{-1}~,
\end{eqnarray}
and encoded operations that commute with the stabilizer and hence act on the
encoded qudit can be chosen to be
\begin{eqnarray}
\bar Z &=& Z_1 Z_4 Z_7~,\nonumber \\
\bar X &=& X_1 X_2 X_3~.
\end{eqnarray}
In the $d\rightarrow \infty$ limit, we obtain a code that is the
simultaneous eigenspace of eight commuting operators acting on nine particles, which are
\begin{eqnarray}
&q_1 - q_2~, q_2 - q_3~, q_4 - q_5~ , q_5 - q_6~ , q_7 - q_8, q_8 - q_9~,\nonumber\\
&(p_1 + p_2 + p_3) - (p_4 + p_5 + p_6)~,\nonumber\\
&(p_4 + p_5 + p_6) - (p_7 +p_8 + p_9)~.
\end{eqnarray}
Logical operators that act in the code space are
\begin{eqnarray}
\bar q &=& q_1 + q_4 + q_7~,\nonumber \\
\bar p &=& p_1 + p_2 + p_3~.
\end{eqnarray}
This code is designed to protect against errors in which one of the particles makes a large
jump in $q$ or $p$ (or both) while the others hold still. But it provides little
protection against small diffusive motions of all the particles, which allow $\bar q$ and $\bar p$ to drift.

Entanglement purification protocols for continuous variable systems have also been proposed --- good entangled states can be distilled from noisy entangled states via a protocol that requires two-way classical communication \cite{plenio,zoller}. These purification protocols work well against certain sorts of errors, but their reliance on two-way classical communication makes them inadequate for accurately preserving unknown states in an imperfect quantum memory, or for robust quantum computation.

Returning to qudits, let us consider an example of a quantum code that can
protect against small shifts in both amplitude and phase, but not against large
shifts.  It is already interesting to discuss the case of a system consisting of a single qudit, but where the dimension $n$ of the encoded system is (of course) less than $d$.  For example, a qubit ($n=2$) can be
encoded in a system with dimension $d=18$, and protected against shifts by one
unit in the amplitude or phase of the qudit; that is, against errors of the
form $X^a Z^b$ where $|a|, |b| \leq 1$.  The stabilizer of this code is generated
by the two operators
\begin{equation}
X^6, \quad Z^6~,
\end{equation}
and the commutation relations of the Pauli operators with these generators are
\begin{eqnarray}
(X^a Z^b)\cdot  X^6 &=& \omega^{6b}~ X^6 \cdot (X^a Z^b)~,\nonumber \\
(X^a Z^b)\cdot  Z^6 &=& \bar\omega^{6a} ~ Z^6 \cdot (X^a Z^b)~.
\end{eqnarray}
Therefore, a Pauli operator commutes with the stabilizer only if $a$ and 
$b$ are both multiples of $3 = 18/6$; this property ensures that the code can
correct single shifts in both amplitude and phase. Logical operators acting on
the encoded qudit are
\begin{equation}
\bar X = X^3~,\quad  \bar Z = Z^3~,
\end{equation}
which evidently commute with the stabilizer and are not contained in it.

Since the codewords are eigenstates of $Z^6$ with eigenvalue one, the only allowed values of $j$ are multiples of three. And since there are also eigenstates of $X^6$ with eigenvalue one, the codewords are invariant under a shift in $j$ by six units. A basis for the two-dimensional code space is
\begin{eqnarray}
|\bar 0\rangle = {1\over\sqrt{3}}\left(|0\rangle +|6\rangle + |12\rangle\right)~,\nonumber\\
|\bar 1\rangle = {1\over \sqrt{3}}\left(|3\rangle + |9\rangle + |15\rangle\right)~.
\end{eqnarray}
If an amplitude error occurs that shifts $j$ by $\pm 1$, the error can be diagnosed by measuring the stabilizer generator $Z^6$, which reveals the value of $j$ modulo 3; the error is corrected by adjusting $j$ to the nearest multiple of 3. Phase errors are shifts in the Fourier transformed conjugate basis, and can be corrected similarly.

This code is actually {\em perfect}, meaning that each possible pair of eigenvalues
of the generators $X^6$ and $Z^6$ is a valid syndrome for correcting a shift.
There are  nine possible errors $\{X^a Z^b,~ |a|, |b| \leq 1\}$, and the
Hilbert space of the qudit contains nine copies of the two-dimensional code
space, one corresponding to each possible error.  These ``error spaces'' just barely fit in the
qudit space for $d = 18=9\cdot 2$.

Similar perfect codes can be constructed that protect against larger shifts.  For $d = r_1
r_2 n$, consider the stabilizer generators
\begin{equation}
X^{r_{1} n}, \quad Z^{r_{2}n}~.
\end{equation}
There is an encoded {\em qunit}, acted on by logical operators
\begin{eqnarray}
\bar X &=& X^{r_{1}}~,\nonumber \\
\bar Z &=& Z^{r_{2}}~,
\end{eqnarray}
which evidently commute with the stabilizer and satisfy
\begin{equation}
\bar Z\bar X= \omega^{r_{1}r_{2}} \bar X \bar Z = e^{2\pi i/n}
\bar X \bar Z~.
\end{equation} 
The commutation relations of the Pauli operators with the
generators are
\begin{eqnarray}
(X^a Z^b) \cdot X^{r_1 n} &=& \omega^{r_{1}nb} ~X^{r_{1}n} \cdot (X^a Z^b)\nonumber \\
&=& e^{2\pi ib/r_2} ~X^{r_{1}n} \cdot (X^a Z^b)~,\nonumber \\
(X^a Z^b)\cdot  Z^{r_{2}n} &=& \bar\omega^{r_{2}na}~ Z^{r_{2}n}\cdot (X^a Z^b)\nonumber \\
&=& e^{-2\pi i a/r_1}~ Z^{r_{2}n} \cdot (X^a Z^b)~.
\end{eqnarray}
The phases are trivial only if $a$ is an integer multiple of $r_1$ and $b$ an integer
multiple of $r_2$.  Therefore, this code can correct all shifts with
\begin{eqnarray}
|a| &<& \frac{r_1}{2} ~,\nonumber \\
|b| &<& \frac{r_2}{2} ~.
\end{eqnarray}
The number of possible error syndromes is $r_1 r_2=d/n$, so the code is perfect.

Expressed in terms of $Z$ eigenstates, the codewords contain only values of $j$ that are multiples of $r_1$ (since $Z^{r_2n}=1$), and are invariant under a shift of $j$ by $r_1 n$ (since $X^{r_1 n}=1$). Hence a  basis for the $n$-dimensional code subspace is
\begin{eqnarray}
&|\bar 0\rangle &= {1\over\sqrt{r_2}}\left(|0\rangle +|nr_1\rangle +\dots +|(r_2-1)nr_1\rangle\right)~,\nonumber\\
&|\bar 1\rangle &= {1\over \sqrt{r_2}}\left(|r_1\rangle + \dots +|((r_2-1)n+1)r_1\rangle\right)~,\nonumber\\
&&\qquad\qquad\cdot\nonumber\\
&&\qquad\qquad\cdot\nonumber\\
&|\bar n-\bar 1\rangle &={1\over \sqrt{r_2}}\left(|(n-1)r_1\rangle + \dots +|(r_2n-1)r_1\rangle\right)~.
\end{eqnarray}
If the states undergo an amplitude shift, the value of $j$ modulo $r_1$ is determined by measuring the stabilizer generator $Z^{r_2 n}$, and the shift can be corrected by adjusting $j$ to the nearest multiple of $r_1$. The codewords have a similar form in the Fourier transformed conjugate basis (the basis of $X$ eigenstates), but with $r_1$ and $r_2$ interchanged. Therefore, amplitude shifts by less than $r_1/2$ and phase shifts by less than $r_2/2$ can be corrected.

\section{A qudit in a Landau level}
\label{sec:landau}
A single electron in a uniform magnetic field in two dimensions provides an enlightening realization of our codes. General translations in a magnetic field are noncommuting, since an electron transported around a closed path acquires an Aharonov-Bohm phase $e^{ie\Phi}$, where $\Phi$ is the magnetic flux enclosed by the path. Two translations $T$ and $S$ commute only if the operator $TST^{-1} S^{-1}$ translates an electron around a path that encloses a flux $\Phi=k\Phi_0$, where $\Phi_0=2\pi/e$ is the flux quantum and $k$ is an integer.

Translations commute with the Hamiltonian $H$, and two translations $T_1$ and $T_2$ form a maximally commuting set if they generate a lattice that has a unit cell enclosing one quantum of flux. Simultaneously diagonalizing $H$, $T_1$ and $T_2$, we obtain a Landau level of degenerate energy eigenstates, one state corresponding to each quantum of magnetic flux. Then $T_1$ and $T_2^n$ are the stabilizer generators of a code, where $\bar Z= T_1^{1/n}$ and $
\bar X=T_2$ are the logical operators on a code space of dimension $n$. 

Suppose the system is in a periodically identified box (a torus), so that $T_1^{r_1}= (T_2^n)^{r_2}=1$ are translations around the cycles of the torus. The number of flux quanta through the torus, and hence the degeneracy of the Landau level, is $nr_1 r_2$. The code, then, embeds an $n$ dimensional system in a system of dimension $d=r_1 r_2 n$.

In this situation, the logical operations $\bar X$ and $\bar Z$ can be visualized as translations of the torus in two different directions; the stabilizer generator $\bar X^n$ is a translation by a fraction $1/r_2$ of the length of the torus in one direction, and the stabilizer generator  $\bar Z^n$ is a translation by $1/r_1$ of the length in the other direction.  Therefore, for any state in the code space, the wave function of the electron in a cell containing $n$ flux quanta is periodically repeated altogether $r_1 r_2$ times to fill the entire torus. Our code can be regarded as a novel kind of ``quantum repetition code'' -- identical ``copies'' of the wave function  are stored in each of $r_1r_2$ cells. But of course there is only one electron, so if we detect the electron in one cell its state is destroyed in all the cells. 

This picture of the state encoded in a Landau level cautions us about the restrictions on the type of error model that the code can fend off successfully. If the environment strongly probes one of the cells and detects nothing, the wave function is suppressed in that cell. This causes a $\bar X$ error in the encoded state with a probability of about $1/2 r_2$, and a $\bar Z$ error with a probability of about $1/2 r_1$. The code is more effective if the typical errors gently deform the state in each cell, rather than strongly deforming it in one cell.

\section{Continuous variable codes for a single oscillator}
\label{sec:sing_osc}
Formally, we can construct quantum codes for systems described by continuous quantum variables by considering the large-$d$ limit of the shift-resistant codes described in \S\ref{sec:shift}. We might have hoped to increase $d$ to infinity while holding  $r_1/d$ and $r_2/d$ fixed, maintaining the ability to correct shifts in both amplitude and phase that are a fixed fraction of the ranges of the qudit. However, since the perfect codes satisfy 
\begin{equation}
\frac{r_1}{d} = \frac{1}{nr_2} ~, \quad \frac{r_2}{d} = \frac{1}{nr_1}~,
\end{equation}
this is not possible. Nonetheless, interesting codes can be obtained as the amplitude and phase of the qudit approach the position $q$ and momentum $p$ of a particle -- we can hold fixed the size of the shifts $\Delta q$ and $\Delta p$ that can be corrected, as the ranges of $q$ and $p$ become unbounded. 

Another option is to take
$d \rightarrow \infty$ with $r_1/d\equiv \frac{1}{m}$ fixed and $r_2 = m/n$
fixed, obtaining a {\it rotor} $Z = e^{i\theta}$ (or a particle in a periodically identified finite box) that can be protected against
finite shifts in both the orientation $\theta$ of the rotor and its (quantized) angular momentum $L$.  The
stabilizer of this code is generated by
\begin{eqnarray}
Z^{r_{2}n} &\rightarrow& e^{i\theta m}~,\nonumber \\
X^{r_{1}n} &=& X^{d/r_2} \rightarrow e^{-2\pi iL(n/m)}
\end{eqnarray}
and the logical operations are
\begin{eqnarray}
\bar Z &=& e^{i\theta m/n}\nonumber \\
\bar X &=& e^{-2\pi iL/m}
\end{eqnarray}
Since $\bar X$ shifts the value of $\theta$ by $2\pi/m$, and $\bar Z$ shifts
the value of $L $ by $m/n = r_2$, this code can correct shifts in $\theta$
with $\Delta\theta < \pi/m$ and shifts in $L$ with $|\Delta L| < m/2n$.

Alternatively, we can consider a limit in which $r_1$ and $r_2$ both become
large. We may write $r_1= \alpha/\varepsilon$ and $r_2= 1/n\alpha\varepsilon$, where $d=nr_1r_2=1/\varepsilon^2$, obtaining a code with stabilizer generators
\begin{eqnarray}
\label{alpha_code}
Z^{r_{2}n} &\rightarrow& \left(e^{2\pi i q\varepsilon}\right)^{(1/\alpha\varepsilon)}= e^{2\pi iq/\alpha}~,\nonumber \\
X^{r_{1}n} &\rightarrow& \left(e^{-ip\varepsilon}\right)^{(n\alpha/\varepsilon)}= e^{-inp\alpha}~,
\end{eqnarray}
and logical operations
\begin{equation}
\bar Z = e^{2\pi iq/n\alpha}~,\quad  \bar X = e^{-ip\alpha}~,
\end{equation}
where $\alpha$ is an arbitrary real number.
Using the identity $e^Ae^B=e^{[A,B]}e^Be^A$ (which holds if $A$ and $B$ commute with their commutator) and the canonical commutation relation $[q,p]=i$, we verify that
\begin{equation}
\bar Z\bar X = \omega \bar X \bar Z~, \quad \omega=e^{2\pi i/n}~.
\end{equation}
Since $\bar X$ translates $q$ by $\alpha$ and $\bar Z$ translates $p$ by
$2\pi/n\alpha$, the code protects against shifts with
\begin{eqnarray}
|\Delta q| &<& \frac{\alpha}{2}~,\nonumber \\
|\Delta p| &<& \frac{\pi}{n\alpha}~.
\end{eqnarray}
Note that the shifts in momentum and position that the code can correct obey the condition
\begin{equation}
\label{delta}
\Delta p \Delta q < {\pi\over 2n}\hbar~.
\end{equation}
In typical situations, errors in $q$ and $p$ are of comparable magnitude, and it is best to choose $\alpha= \sqrt{2\pi/n}$ so that
\begin{equation}
\bar Z= \exp\left(iq\sqrt{2\pi\over n} \right)~,\quad \bar X= \exp\left (-ip\sqrt{2\pi\over n} \right)~.
\end{equation}

\begin{figure}
\begin{center}
\leavevmode
\epsfxsize=3in
\epsfbox{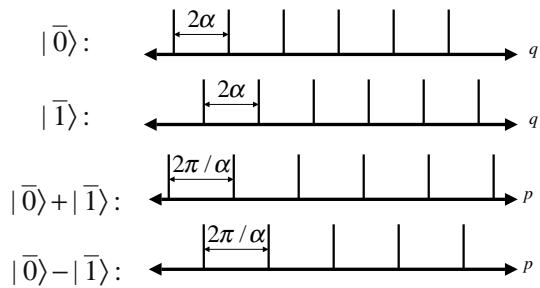}
\end{center}
\caption{Codewords of the $n=2$ code. The states $|\bar 0\rangle$, $|\bar 1\rangle$ are superpositions of $q$ eigenstates, periodically spaced with period $2\alpha$; the two basis states differ by a displacement in $q$ by $\alpha$. The states $(|0\rangle\pm |1\rangle)/\sqrt{2}$ are superpositions of $p$ eigenstates, periodically spaced with period $2\pi/\alpha$; the two basis states differ by a displacement in $p$ by $\pi/\alpha$.}
\label{fig_codewords}
\end{figure}

Formally, the codewords are coherent superpositions of infinitely squeezed
states, {\it e.g.} (up to normalization)
\begin{eqnarray}
|\bar Z = \omega^j\rangle&=& \sum_{s = - \infty}^{\infty} |q = \alpha (j +
ns)\rangle~,\nonumber \\
|\bar X = \bar \omega^j\rangle&=& \sum_{s = - \infty}^{\infty} |p =
\frac{2\pi}{n\alpha} (j + ns)\rangle~.
\end{eqnarray}
(See Fig.~\ref{fig_codewords}.) Of course, realistic codewords will be normalizable finitely squeezed states, rather than nonnormalizable infinitely squeezed states. But squeezing in at least one of $p$ and $q$ is required to comfortably fulfill the condition eq.~(\ref{delta}).

The Wigner function associated with the codeword wave function $\psi^{(j)}(q)\equiv \langle q|\bar Z=\omega^j\rangle$ is a sum of delta functions positioned at the sites of a lattice in phase space, where three quarters of the delta functions are positive and one quarter are negative. Explicitly, we have 
\begin{eqnarray}
W^{(j)}(q,p)&\equiv& {1\over 2\pi}\int_{-\infty}^{\infty}dx~e^{ipx}
\psi^{(j)}(q+x/2)^*\psi^{(j)}(q-x/2)\nonumber\\
&\propto& \sum_{s,t=-\infty}^{\infty}(-1)^{st} \delta\left(p- {\pi \over n\alpha}\cdot s\right)
\nonumber\\
& & \times ~\delta\left(q-\alpha j -{n \alpha\over 2}\cdot t\right)
~;
\end{eqnarray}
the delta functions are negative on the sublattice with $s,t$ odd. If we integrate over $p$, the oscillating sign causes the terms with odd $t$ to cancel in the sum over $s$, and the surviving positive delta functions have support at $q= (n\cdot{\rm integer}+ j)\alpha$. If we integrate over $q$, the terms with odd $s$ cancel in the sum over $t$, and the surviving positive delta functions have support at $p=(2\pi/n\alpha)\cdot {\rm integer}$.
Wigner functions for the $\bar X$ eigenstates are similar, but with the roles of $q$ and $p$ interchanged.

It is also of interest to express the encoded states in terms of the basis of coherent states. Consider for example the encoded state with $\bar X=1$, which is the unique simultaneous eigenstate with eigenvalue one of the operators $e^{2\pi i q/\alpha}$ and $e^{-ip\alpha}$. In fact starting with any state $|\psi\rangle$, we can construct the encoded state (up to normalization) as 
\begin{eqnarray}
&\left(\sum_{s=-\infty}^\infty e^{-isp\alpha}\right)\cdot \left(\sum_{t=-\infty}^\infty e^{2\pi i t q/\alpha}\right) |\psi\rangle \nonumber\\
& =\sum_{s,t}\exp \left[i \left(-sp\alpha + 2\pi  t q/\alpha + \pi st\right) \right]|\psi\rangle~.
\end{eqnarray}
In particular, if we choose $|\psi\rangle$ to be the ground state $|0\rangle$ of the oscillator, then the operator $\sum_{s,t} \exp \left[i  \left(-sp\alpha + 2\pi  t q/\alpha +\pi st\right)\right]$ displaces it to a coherent state centered at the point $(q,p) = (s\alpha,2\pi t/\alpha )$ in the quadrature plane. Thus the encoded state is an equally weighted superposition of coherent states, with centers chosen from the sites of a lattice in the quadrature plane whose unit cell has area $2\pi$. Since the coherent states are overcomplete the expansion is not unique; indeed, if we choose $|\psi\rangle$ to be a coherent state rather than the vacuum, then the lattice is rigidly translated, but the encoded state remains invariant.

We can envision the stabilizer of the code as a lattice of translations in phase space that preserve the code words, the lattice generated by the translations $e^{2\pi iq/\alpha}$ and $e^{-inp \alpha}$. In fact, this lattice need not be rectangular -- we can encode an $n$-dimensional system in the Hilbert space of a single oscillator by choosing {\em any} two variables $Q$ and $P$ that satisfy the canonical commutation relation $[Q,P]=i$, and constructing the code space as the simultaneous eigenstate of $e^{2\pi i Q}$ and $e^{-i n P}$. The unit cell of the lattice has area $2\pi \hbar n$, in keeping with the principle that each quantum state ``occupies'' an area $2\pi\hbar$ in the phase space of a system with one continuous degree of freedom.

\section{Finite squeezing}
\label{sec:finite}

Strictly speaking, our codewords are nonnormalizable states, infinitely squeezed in both $q$ and $p$. In practice, we will have to work with approximate codewords that will be finitely squeezed normalizable states. We need to consider how using such approximate codewords will affect the probability of error.

\begin{figure}
\begin{center}
\leavevmode
\epsfxsize=3in
\epsfbox{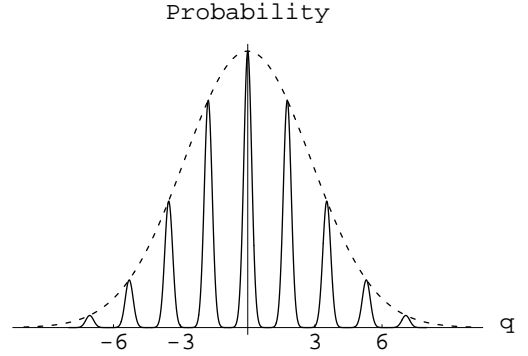}
\end{center}
\caption{Probability distribution in position space $P(q)= {1\over 2}|\langle q|(|\tilde 0\rangle + |\tilde 1\rangle)|^2$ for an approximate codeword with $\Delta=\kappa=.25$. The dashed line is the distribution's Gaussian envelope.}
\label{fig_envelope}
\end{figure}

We will replace a position eigenstate $\delta(0)$ by a normalized Gaussian of width $\Delta$ centered at the origin,
\begin{eqnarray}
|\psi_0\rangle &=& \int_{-\infty}^{\infty} {dq\over (\pi\Delta^2)^{1/4}} e^{-{1\over 2}q^2/\Delta^2}|q\rangle \nonumber\\
&=&  \int_{-\infty}^{\infty} {dp\over (\pi/\Delta^2)^{1/4}} e^{-{1\over 2}\Delta^2p^2}|p\rangle~.
\end{eqnarray}
A codeword, formally a coherent superposition of an infinite number of $\delta$-functions, becomes a sum of Gaussians weighted by a Gaussian envelope function of width $\kappa^{-1}$; in the special case of a two-dimensional code space, the approximate codewords become
\begin{eqnarray}
|\tilde 0\rangle &=&  N_0\sum_{s=-\infty}^{\infty}e^{-{1\over 2} \kappa^2 (2s\alpha)^2}T(2s\alpha)|\psi_0\rangle~,\nonumber\\
|\tilde 1\rangle &=&  N_1\sum_{s=-\infty}^{\infty}e^{-{1\over 2} \kappa^2 [(2s+1)\alpha)]^2}T[(2s+1)\alpha]|\psi_0\rangle~,
\end{eqnarray}
where $T(a)$ translates $q$ by $a$, $N_{0,1}$ are normalization factors, and we use {\it e.g.} $|\tilde 0\rangle$ rather than $|\bar 0\rangle$ to denote the approximate codeword. We will assume that $\kappa\alpha$ and $\Delta/\alpha$ are small compared to one, so that $N_0\approx N_1\approx (4\kappa^2\alpha^2/\pi)^{1/4}$; then in momentum space, the approximate codeword becomes {\it e.g.}, 
\begin{eqnarray}
& (|\tilde 0 \rangle + |\tilde 1\rangle)/\sqrt{2}& ~\approx  \left(\kappa^2\alpha^2\over\pi\right)^{1/4} \int_{-\infty}^{\infty} {dp\over (\pi/\Delta^2)^{1/4}} e^{-{1\over 2}\Delta^2p^2}\nonumber\\
& &\times \sum_{s=-\infty}^{\infty}e^{-{1\over 2} \kappa^2 (s\alpha)^2}e^{ip(\alpha s)}|p\rangle~.
\end{eqnarray}
By applying the Poisson summation formula, \begin{equation}
\sum_{m=-\infty}^{\infty}e^{-\pi a(m-b)^2}=(a)^{-1/2}\sum_{s=-\infty}^{\infty}e^{-\pi s^2/a}e^{2\pi i sb}~,
\end{equation}
this approximate codeword can be rewritten as
\begin{eqnarray}
& (|\tilde 0 \rangle + &|\tilde 1\rangle)/\sqrt{2} \approx  \left(\kappa^2\alpha^2\over\pi\right)^{1/4} \int_{-\infty}^{\infty} {dp\over (\pi/\Delta^2)^{1/4}} e^{-{1\over 2}\Delta^2p^2}\nonumber\\
& &\times {\sqrt{2\pi}\over \kappa \alpha}\sum_{m=-\infty}^{\infty}\exp\left[-{1\over 2} \left(p-{2\pi\over\alpha}m\right)^2/\kappa^2\right]|p\rangle~\nonumber\\
& &= \int_{-\infty}^{\infty}{dp\over (\pi\kappa^2)^{1/4}}\cdot \left({4\pi\Delta^2\over \alpha^2}\right)^{1/4}\sum_{m=-\infty}^{\infty}e^{-{1\over 2}\Delta^2p^2}
\nonumber\\
& & \times\exp\left[-{1\over 2} \left(p-{2\pi\over\alpha}m\right)^2/\kappa^2\right]|p\rangle~,
\end{eqnarray}
again a superposition of Gaussians weighted by a Gaussian envelope. (See Fig.~\ref{fig_envelope}.)

The approximate codewords $|\tilde 0\rangle, |\tilde 1\rangle$ have a small overlap if $\Delta$ is small compared to $\alpha$ and $\kappa$ is small compared to $\pi/\alpha$. For estimating the error probability caused by the overlap, let's consider the special case where $q$ and $p$ are treated symmetrically, $\alpha=\sqrt{\pi}$ and $\kappa=\Delta$; then 
\begin{equation}
|\langle q|\tilde 0\rangle|^2\approx {2\over\sqrt{\pi}}\sum_{s=-\infty}^{\infty}e^{-4\pi \Delta^2s^2}\exp[-(q-2s\sqrt{\pi} )^2/\Delta^2]~
\end{equation}
and 
\begin{eqnarray}
& &{1\over 2}\left|\langle p|\tilde 0\rangle + \langle p|\tilde 1\rangle\right|^2\nonumber\\
& &\approx {2\over\sqrt{\pi}}\sum_{m=-\infty}^{\infty}e^{-\Delta^2p^2}\exp[-(p-2m\sqrt{\pi} )^2/\Delta^2]~.
\end{eqnarray}
To perform error recovery, we measure the value of $q$ and $p$ modulo $\sqrt{\pi}$ and then correct for the observed shift. In the state $|\tilde 0\rangle$, the probability of failure is the probability that $q$ is closer to an odd multiple of $\sqrt{\pi}$ than an even multiple, and in the state $(|\tilde 0\rangle + |\tilde 1\rangle)/\sqrt{2}$, the error probability is the probability that $p$ is closer to an odd multiple of $\sqrt{\pi}$ than an even multiple.  For both the amplitude and phase errors, the intrinsic error probability arising from the imperfections of the approximate codewords becomes exponentially small for small $\Delta$. Using  the asymptotic expansion of the error function,
\begin{equation}
\int_x^\infty dt~e^{-t^2}= \left({1\over 2x}\right)e^{-x^2}\left(1 - O(1/x^2)\right)~,
\end{equation}
we may estimate the error probability by summing the contributions from the tails of all the Gaussians, obtaining
\begin{eqnarray}
{\rm Error ~ Prob} &< & {2\over {\sqrt{\pi}}}
\left(\sum_{n=-\infty}^\infty e^{-4\pi\Delta^2n^2}\right)\cdot 2 \int_{\sqrt\pi/2}^\infty dq ~e^{-q^2/\Delta^2}\nonumber\\
& \sim &{2\over\sqrt{\pi}}\cdot {1\over 2\Delta}\cdot 2\Delta\cdot {\Delta\over \sqrt{\pi}}e^{-\pi/4\Delta^2}\nonumber\\
&= &{2\Delta\over \pi}e^{-\pi/4\Delta^2}~.
\end{eqnarray}

This error probability is about $1\%$ for $\Delta\sim .5$, and is already less than $10^{-6}$ for a still modest value $\Delta\sim .25$. Using finitely squeezed approximate codewords does not badly compromise the error-correcting power of the code, since a gentle spreading in $p$ and $q$ is just the kind of error the code is intended to cope with.

The mean photon number of a finitely squeezed approximate codeword is 
\begin{equation}
\langle a^\dagger a \rangle +1/2 = {1\over 2}\langle p^2 + q^2\rangle \approx \Delta^{-2}
\end{equation}
for small $\Delta$. Therefore, an error probability of order $10^{-6}$ can be achieved with Gaussian approximate codewords that have mean photon number of about $(.25)^{-2}\sim 16$.

More generally, a finitely squeezed codeword $|\psi\rangle$ can be regarded as a perfect codeword $|\xi\rangle$ that has
undergone an error; we may write
\begin{equation}\label{error_wave_function}
|\psi\rangle\, =\,
\int du\, dv\, \eta(u,v)\, e^{i(-up+vq)} |\xi\rangle,
\end{equation}
where $\eta(u,v)$ is an error
``wave function.'' In the special case of a Gaussian finitely squeezed codeword, we have
\begin{equation}\label{Gaussian_error}
\eta(u,v)\, =\, \frac{1}{\sqrt{\pi\kappa\Delta}}
\exp\Bigl(-\frac{1}{2}(u^2/\Delta^2+v^2/\kappa^2)\Bigr)~,
\end{equation}
where $\Delta$ and $\kappa$ are the squeezing parameters defined above. 

If $\eta(u,v)$
vanishes for $|u|>\alpha/2$ or $|v|>\pi/(n\alpha)$, then the error is correctable. In this case, the interpretation of
$\eta(u,v)$ as a wave function has a precise meaning, since there is an unambiguous decomposition of a state into codeword and error. Indeed, if $|\xi_1\rangle$,  $|\xi_2\rangle$ are perfect codewords and 
$|\psi_1\rangle$, $|\psi_2\rangle$ are the corresponding finitely squeezed codewords with error wave functions $\eta_1$,
$\eta_2$, then
\begin{equation}
\langle\psi_1|\psi_2\rangle\,=\,
\langle\xi_1|\xi_2\rangle\, \langle\eta_1|\eta_2\rangle~,
\end{equation}
where
\begin{equation}
\langle\eta_1|\eta_2\rangle=\int du\,dv ~\eta_1(u,v)^*\, \eta_2(u,v)~.
\end{equation}

\section{Continuous variable codes for many oscillators}
\label{sec:many_osc}
The continuous variable codes described in \S \ref{sec:sing_osc} are based on simple lattices in the two-dimensional phase space of a single particle. We can construct more sophisticated codes from lattices in the $(2N)$-dimensional phase space of $N$ particles.  Then codes of higher quality can be constructed that take advantage of efficient packings of spheres in higher dimensions.

For a system of $N$ oscillators, a tensor product of Pauli operators can be expressed in terms of the canonical variables $q_i$ and $p_i$ as 
\begin{equation}
\label{general_pauli}
U_{\alpha \beta} =\exp \left[i \sqrt{2\pi}\left(\sum_{i=1}^N \alpha_i p_i +
\beta_i q_i\right)\right]~,
\end{equation}
where the $\alpha_i$'s and $\beta_i$'s are real numbers. (In this setting, the Pauli operators are sometimes called ``Weyl operators.'')
Two such operators commute up to a phase: 
\begin{eqnarray}
U_{\alpha \beta} U_{\alpha' \beta'} =e^{
2\pi i[\omega(\alpha\beta,\alpha'\beta')]}
U_{\alpha' \beta'}U_{\alpha\beta}~,
\end{eqnarray}
where
\begin{equation}
\omega(\alpha\beta,\alpha'\beta')\equiv \alpha\cdot \beta' - \alpha' \cdot \beta
\end{equation}
is the symplectic form. Thus two Pauli operators commute if and only if their symplectic form is an integer.

Now a general continuous variable stabilizer code is the simultaneous eigenspace of commuting Pauli operators, the code's stabilizer generators. If the continuous variable phase space is $2N$-dimensional and the code space is a finite-dimensional Hilbert space, then there must be $2N$ independent generators. The elements of the stabilizer group are in one-to-one correspondence with the points of a lattice ${\cal L}$ in phase space, via the relation
\begin{equation}
U(k_1,k_2,\dots k_{2N})= \exp \left[i \sqrt{2\pi}\left(\sum_{a=1}^{2N}k_a v_a\right)\right]~.
\end{equation}
Here $\{v_a, a=1,2,\dots, 2N\}$ are the basis vectors of the lattice (each a linear combination of $q$'s and $p$'s), the $k_a$'s are arbitrary integers, and  $U(k_1,k_2,\dots k_{2N})$ is the corresponding element of the stabilizer. For the stabilizer group to be Abelian, the symplectic inner product of any pair of basis vectors must be an integer; that is, the antisymmetric $2N \times 2N$ matrix
\begin{equation}
\label{lattice_a}
A_{ab}= \omega (v_a,v_b)
\end{equation}
has integral entries.
The lattice ${\cal L}$ has a $2N\times 2N$ generator matrix $M$ whose rows are the basis vectors,
\begin{equation}
M=\pmatrix{v_1\cr v_2\cr \cdot\cr \cdot\cr v_{2N}\cr}~.
\end{equation}
In terms of $M$, the matrix $A$ can be expressed as 
\begin{equation}
A= M\omega M^T~,
\end{equation}
where $\omega$ denotes the $2N\times 2N$ matrix 
\begin{equation}
\omega= \pmatrix{0 & I \cr -I & 0 \cr}~,
\end{equation}
and $I$ is the $N\times N$ identity matrix.

The generator matrix of a lattice is not unique. The replacement
\begin{equation}
M\to M'=RM
\end{equation}
leaves the lattice unmodified, where $R$ is an invertible integral matrix with determinant $\pm 1$ (whose inverse is also integral). Under this replacement, the matrix $A$  changes according to
\begin{equation}
A\to A'=RAR^T~.
\end{equation}
By Gaussian elimination, an $R$ can be constructed such that the antisymmetric matrix $A$ is transformed to
\begin{equation}
\label{standard_a}
A'=\pmatrix{0&D\cr -D&0\cr}~,
\end{equation}
where $D$ is a positive diagonal $N\times N$ matrix. 

There are also Pauli operators that provide a basis for the operations acting on the code subspace -- these are the Pauli operators that commute with the stabilizer but are not contained in the stabilizer. The operators that commute with the stabilizer themselves form a lattice ${\cal L}^\perp$ that is dual (in the symplectic form) to the stabilizer lattice. The basis vectors of this lattice can be chosen to be $\{u_b, b=1,2,3,\dots, 2N\}$ such that
\begin{equation}
\label{dual_basis}
\omega(u_a,v_b)= \delta_{ab}~;
\end{equation}
then the generator matrix 
\begin{equation}
M^\perp=\pmatrix{u_1\cr u_2\cr \cdot\cr \cdot\cr u_{2N}\cr}~
\end{equation}
of ${\cal L}^\perp$ has the property 
\begin{equation}
M^\perp\omega M^T= I~.
\end{equation}

It follows from eq.~(\ref{lattice_a}) and eq.~(\ref{dual_basis}) that the ${\cal L}$ basis vectors can be expanded in terms of the ${\cal L}^\perp$ basis vectors as
\begin{equation}
v_a= \sum_b~A_{ab}u_b~,
\end{equation}
or
\begin{equation}
M= AM^\perp~,
\end{equation}
and hence that 
\begin{equation}
\omega(u_a,u_b)=\left(A\right)^{-1}_{ba}~,
\end{equation}
or
\begin{equation}
M^\perp\omega \left(M^\perp\right)^T= \left(A^{-1}\right)^T~.
\end{equation}
If the lattice basis vectors are chosen so that $A$ has the standard form eq.~(\ref{standard_a}), then 
\begin{equation}
\left(A^{-1}\right)^T=\pmatrix{0&D^{-1}\cr -D^{-1}&0\cr}~.
\end{equation}
In the special case of a self-dual lattice, corresponding to a code with a one-dimensional code space, both $A$ and $A^{-1}$ must be integral; hence $D=D^{-1}$ and the standard form of $A$ is
\begin{equation}
A=\pmatrix{0&I\cr -I&0\cr}=\omega~.
\end{equation}

Since the code subspace is invariant under the translations in ${\cal L}$, we can think of the encoded information as residing on a torus, the unit cell of ${\cal L}$. The encoded Pauli operators $\{\bar X^a \bar Z^b\}$ are a lattice of translations on this torus, corresponding to the coset space ${\cal L}^\perp/{\cal L}$. The number of encoded Pauli operators is the ratio of the volume of the unit cell of ${\cal L}$ to the volume of the unit cell of ${\cal L}^\perp$, namely the determinant of $A$, which is therefore the square of the dimension of the Hilbert space of the code. Thus the dimension of the code space is 
\begin{equation}
n=|{\rm Pf}~A|= {\rm det}~D~,
\end{equation}
where ${\rm Pf}~A$ denotes the {\em Pfaffian}, the square root of the determinant of the antisymmetric matrix $A$.

The stabilizer lattice unit cell has volume $|{\rm Pf}~A|$ in units with $h=2\pi\hbar=1$, and the unit cell of the lattice of encoded operations has volume $|{\rm Pf}~ A|^{-1}$ in these units. So the code fits an $n$-dimensional code space into $n$ units of phase space volume, as expected.

Codes of the CSS type (those analogous to the binary quantum codes first constructed by Calderbank and Shor \cite{cal_shor} and by Steane \cite{steane}) are constructed by choosing one lattice ${\cal L}_q$ describing stabilizer generators that are linear combinations of the $q$'s, and another lattice ${\cal L}_p\subset {\cal L}_q^\perp$ describing stabilizer generators that are linear combinations of the $p$'s. (Here ${\cal L}_q^\perp$ denotes the {\em Euclidean} dual of the lattice ${\cal L}_q$.)
The generator matrix of a CSS code has the form
\begin{equation}
M=\pmatrix{M_q & 0\cr 0 & M_p}~,
\end{equation}
where $M_q$ and $M_p$ are $N\times N$ matrices, and the integral matrix $A$ has the form
\begin{equation}
A = \pmatrix{0 & M_qM_p^T \cr
-M_pM_q^T & 0\cr}
\end{equation}

\begin{figure}
\begin{center}
\leavevmode
\epsfxsize=3in
\epsfbox{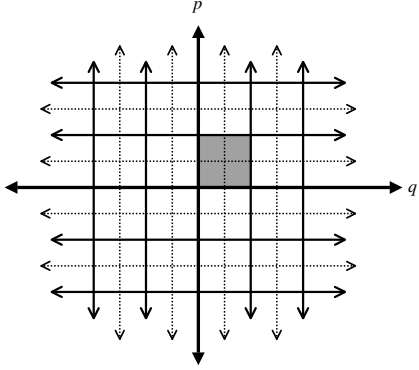}
\end{center}
\caption{The stabilizer lattice and its dual for an $n=2$ code of a single oscillator. Solid lines indicate the stabilizer lattice; solid and dotted lines together comprise the dual lattice. In units of $(2\pi\hbar)^2$, the unit cell of the stabilizer lattice (shaded) has area 2, and the unit cell of its dual has area 1/2.}
\label{fig_lattice}
\end{figure}

For single-oscillator codes described in \S\ref{sec:sing_osc}, $A$ is the $2\times 2$ matrix
\begin{equation}
A = \pmatrix{0 & n \cr
-n & 0\cr}~,
\end{equation}
where $n$ is the code's dimension. For a single-oscillator CSS code, the lattice is rectangular, as shown in Fig.~\ref{fig_lattice}.

The closest packing of circles in two dimensions is achieved by the hexagonal lattice. The generator matrix for a hexagonally encoded qunit can be chosen to be
\begin{equation}
M=\left({2\over\sqrt{3}}\cdot n\right)^{1/2}\pmatrix{1 & 0\cr {1/2 }& {\sqrt{3}/2}\cr}~,
\end{equation}
and the dual lattice is generated by
\begin{equation}
M^\perp= {1\over n}\cdot M~.
\end{equation}
The shortest vector of the dual lattice has length $\left(2/n\sqrt{3}\right)^{1/2}$, compared to length $1/\sqrt{n}$ for the square lattice. Therefore the size of the smallest uncorrectable shift is larger for the hexagonal code than for the square lattice code, by the factor $\left(2/\sqrt{3}\right)^{1/2}\approx 1.07457$. 

An important special class of quantum codes for many oscillators are the {\em concatenated codes}. In particular, we can encode a qubit in each of $N$ oscillators using the code of \S\ref{sec:sing_osc}.
Then we can use a binary stabilizer code that encodes $k$ qubits in a block of $N$ oscillators, and protects against arbitrary errors on any $t$ oscillators, where $2t+1$ is the binary code's distance. The concatenated codes have the important advantage that they can protect against a broader class of errors than small diffusive shifts applied to each oscillator -- if most of the oscillators undergo only small shifts in $p$ and $q$, but a few oscillators sustain more extensive damage, then concatenated codes still work effectively. 

For example, there is a binary $[[7,1,3]]$ quantum code, well suited to fault-tolerant processing, that encodes one logical qubit in a block of seven qubits and can protect against heavy damage on any one of the seven \cite{steane_7}. Given seven oscillators, we can encode a qubit in each one that is resistant to quantum diffusion, and then use the $[[7,1,3]]$ block code to protect one logical qubit against severe damage to any one of the oscillators.

For $n\ge 5$, there is a [[5,1,3]] polynomial code \cite{ben-or}, also well suited to fault-tolerant processing, encoding one qunit in a block of 5. (Actually, $[[5,1,3]]$ quantum codes exist for $n<5$ as well \cite{chau,rains}, but these codes are less conducive to fault-tolerant computing.) The larger value of $n$ increases the vulnerability of each qunit to shift errors. Hence, whether the [[7,1,3]] binary code or the [[5,1,3]] should be preferred depends on the relationship of the size of the typical shift errors to the rate of large errors.

\section{Error models}
\label{sec:error}
What sort of errors can be corrected by these codes?  The codes are designed to protect against errors that shift the values of the canonical variables $p$ and $q$. In fact the Pauli operators are a complete basis, so the action of a general superoperator ${\cal E}$ acting on the input density matrix $\rho$ of a single oscillator can be expanded in terms of such shifts, as in
\begin{eqnarray}
{\cal E}(\rho)& = &\int d\alpha d\beta d\alpha' d\beta' ~C(\alpha,\beta;\alpha'\beta') \nonumber\\
&\times & e^{i(\alpha p + \beta q)}
 ~\rho~ e^{-i(\alpha' p + \beta' q)}~.
\end{eqnarray}
If the support of $C(\alpha,\beta;\alpha',\beta')$ is concentrated on sufficiently small values of its arguments, then the input $\rho$ can be recovered with high fidelity.

A useful model of decoherence is the special case of a ``Pauli channel'' in which $C(\alpha,\beta;\alpha',\beta')$ is diagonal and the superoperator can be expressed as 
\begin{equation}
{\cal E}(\rho) = \int d\alpha d\beta \, ~P(\alpha,\beta)~
e^{i(\alpha p + \beta q)}
 ~\rho~ e^{-i(\alpha p + \beta q)}~.
\end{equation}
Since ${\cal E}$ is positive and trace preserving, we infer that $P(\alpha,\beta)\ge 0$ and
\begin{equation}
\int d\alpha d\beta ~P(\alpha,\beta)=1~.
\end{equation}
Thus, we may interpret $P(\alpha,\beta)$ as a probability distribution: the phase space translation
\begin{equation}
(q,p) \to (q - \alpha, p +\beta)
\end{equation}
is applied with probability $P(\alpha,\beta)$.

Weak interactions between an oscillator and its environment drive a diffusive process that can be well modeled by a Pauli channel. If the environment quickly ``forgets'' what it learns about the oscillator, the evolution of the oscillator can be described by a master equation. Over a short time interval $dt$, the shifts applied to the oscillator may be assumed to be small, so that the Pauli operator can be expanded in powers of $\alpha$ and $\beta$. Suppose that the shifts are symmetrically distributed in phase space such that 
\begin{eqnarray}
&& \langle \alpha \rangle = \langle \beta\rangle = 0~,\nonumber\\
&& \langle \alpha^2\rangle = \langle \beta^2\rangle~, \nonumber\\
&& \langle \alpha\beta\rangle = 0~,
\end{eqnarray}
where $\langle \cdot \rangle$ denotes the mean value determined by the probability distribution $P(\alpha,\beta)$. Suppose further that the shifts are diffusive, so that the mean square displacement increases linearly with $dt$; we may write
\begin{equation} 
\langle \alpha^2\rangle = \langle \beta^2\rangle = D dt~,
\end{equation}
where $D$ is a diffusion constant. 
We then obtain
\begin{eqnarray}
\rho(t+dt)& =& \int  d\alpha d\beta\, ~P(\alpha,\beta)~
e^{i(\alpha p + \beta q)}
 \rho~ e^{-i(\alpha p + \beta q)}~ \nonumber\\
& = &\rho(t) + D dt \left( p\rho p -{1\over 2} p^2\rho - {1\over 2}\rho p^2\right)\nonumber\\
&+&  Ddt\left(q\rho q - {1\over 2} q^2\rho - {1\over 2}\rho q^2\right) + O(dt^{3/2})~,
\end{eqnarray}
or 
\begin{equation}
\dot \rho  = -{D\over 2}[p,[p,\rho]] -{D\over 2}[q,[q,\rho]] ~.
\end{equation}
The interpretation of $D$ as a diffusion constant can be confirmed by computing
\begin{equation}
\frac{d}{dt} {\rm tr}~\left(p^2\rho\right)= D = \frac{d}{dt} {\rm tr}~\left(q^2\rho\right)~;
\end{equation}
the mean square values of $p$ and $q$ increase with time as $Dt$.

More generally, the master equation contains a diffusive term determined by the covariance of the distribution $P(\alpha,\beta)$, and perhaps also a nondissipative drift term determined by the mean of $P(\alpha,\beta)$. Our quantum error-correcting codes can successfully suppress decoherence caused by quantum diffusion, if the recovery operation is applied often enough; roughly, the time interval $\Delta t$ between error correction steps should be small compared to the characteristic diffusion time $D^{-1}$.

Interactions with the environment might also damp the amplitude of the oscillator, as described by the master equation
\begin{equation}
\dot \rho= \Gamma\left( a\rho a^\dagger - {1\over 2}a^\dagger a \rho - {1\over 2}\rho a^\dagger a\right)~;
\end{equation}
here $a=(q+ip)/\sqrt{2}$ is the annihilation operator and $\Gamma$ is a decay rate. This master equation cannot be obtained from a Pauli channel, but as for quantum diffusion,  the effects of amplitude damping over short time intervals can be expressed in terms of small phase-space displacements. 

The master equation for amplitude damping can be obtained as the $dt\to 0$ limit of the superoperator
\begin{eqnarray}
\rho(t+dt)&=& {\cal E}(\rho(t))= \left(\sqrt{\Gamma dt}~a\right)\rho(t)
\left(\sqrt{\Gamma dt}~a^\dagger\right)\nonumber\\
&&+\left(I -{\Gamma dt\over 2}a^\dagger a\right)\rho(t)\left(I -{\Gamma dt\over 2}a^\dagger a\right)~.
\end{eqnarray}
For $dt$ small, the annihilation operator can be expanded in terms of Pauli operators as
\begin{eqnarray}
\sqrt{\Gamma dt} ~a& \approx& -{i\over 2}\left(e^{i\sqrt{\Gamma dt/2}~q}-e^{-i\sqrt{\Gamma dt/2}~q}\right)
\nonumber\\
&&+{1\over 2}\left(e^{i\sqrt{\Gamma dt/2}~p}-e^{-i\sqrt{\Gamma dt/2}~p}\right)
\end{eqnarray}
Thus, if the time interval $\Delta t$ between error correction steps is small compared to the damping time $\Gamma^{-1}$, the displacements applied to codewords are small, and error correction will be effective.

Aside from decoherence, we also need to worry about ``unitary errors.'' For example, an inadvertent rotation of the phase of the oscillator induces the unitary transformation
\begin{equation}
U(\theta)\equiv \exp\left( i\theta a^\dagger a \right)
\end{equation}
Like any unitary transformation, this phase rotation can be expanded in terms of Pauli operators. It is convenient to introduce the notation for the phase-space displacement operator
\begin{eqnarray}
D(\gamma)&\equiv& \exp\left( \gamma a - \gamma^* a^\dagger\right)\nonumber\\
&=& \exp \left(i\sqrt{2} \left[({\rm Im}~\gamma)q-({\rm Re}~\gamma)p\right]\right)~,
\end{eqnarray}
where $\gamma$ is a complex number. The displacements satisfy the identity
\begin{equation}
{\rm tr}\left(D(\gamma) D(\eta)^\dagger\right) = \pi \delta^2(\gamma - \eta)~,
\end{equation}
so the operator $U(\theta)$ can be expanded in terms of displacements as
\begin{equation}
U(\theta)= {1\over \pi} \int d^2 \gamma ~u_\theta(\gamma) D(\gamma)~,
\end{equation}
where
\begin{equation}
u_\theta(\gamma) ={\rm tr}\left( U(\theta) D(\gamma)^\dagger \right)~.
\end{equation}
Evaluating the trace in the coherent state basis, we find that
\begin{equation}
u_\theta(\gamma)= {i~e^{i\theta/2}\over 2\sin(\theta/2)}
\exp\left(- {i\over 2}|\gamma|^2\cot (\theta/2)\right)~.
\end{equation}
For small $\theta$, the coefficient
\begin{equation}
u_\theta(\gamma)\approx {i\over \theta}\exp\left(-~{i\over\theta}|\gamma|^2\right)~
\end{equation}
has a rapidly oscillating phase, and can be regarded as a distribution with support concentrated on values of $\gamma$ such that $|\gamma|^2\sim \theta$; indeed, formally
\begin{equation}
\lim_{\theta\to 0}~u_\theta(\gamma)= \pi ~\delta^2(\gamma)~.
\end{equation}
Thus a rotation by a small angle $\theta$ can be accurately expanded in terms of small displacements -- error correction is effective if an oscillator is slightly overrotated or underrotated. 
\section{The Gaussian quantum channel}
\label{sec:gaussian}

At what rate can error-free digital information be conveyed by a noisy continuous signal? In classical information theory, an answer is provided by Shannon's noisy channel coding theorem for the Gaussian channel \cite{cover}. This theorem establishes the capacity that can be attained by a signal with specified average power, for a channel with specified bandwidth and specified Gaussian noise power. The somewhat surprising conclusion is that a nonzero rate can be attained for any nonvanishing value of the average signal power. 

A natural generalization of the Gaussian classical channel is the {\em Gaussian quantum channel}. The Gaussian quantum channel is a Pauli channel: $N$ oscillators are transmitted, and the $q$ and $p$ displacements acting on the oscillators are independent Gaussian random variables with mean 0 and variance $\sigma^2$. A code is an $M$-dimensional subspace of the Hilbert space of the $N$ oscillators, and the rate $R$ of the code (in qubits) is defined as
\begin{equation}
R= {1\over N}\log_2 M~.
\end{equation}
The quantum-information capacity $C_Q$ of the channel is the maximal rate at which quantum information can be transmitted with fidelity arbitrarily close to one.

The need for a constraint on the signal power to define the capacity of the Gaussian classical channel can be understood on dimensional grounds. The classical capacity (in bits) is a dimensionless function of the variance $\sigma^2$, but $\sigma^2$ has dimensions. Another quantity with the same dimensions as $\sigma^2$ is needed to construct a dimensionless variable, and the power fulfills this role. But no power constraint is needed to define the quantum capacity of the quantum channel. The capacity (in qubits) is a function of the dimensionless variable $\hbar/\sigma^2$, where $\hbar$ is Planck's constant.

An upper bound on the quantum capacity of the Gaussian quantum channel was derived by Holevo and Werner \cite{holevo}; they obtained (reverting now to units with $\hbar=1$)
\begin{equation} 
C_Q \le \log_2 (1/\sigma^2)~,
\end{equation}
for $0<\sigma^2 < 1$, and $C_Q=0$ for $\sigma^2\ge 1$.
They also computed the coherent information $I_Q$ of the Gaussian quantum channel, and maximized it over Gaussian signal states, finding \cite{holevo}
\begin{equation}
\label{coherent_info}
\left(I_Q\right)_{\rm max}= \log_2 (1/e\sigma^2)~,
\end{equation}
for $0< \sigma^2 <1/e$ (where $e=2.71828\dots)$. The coherent information is {\em conjectured} to be an attainable rate \cite{lloyd_conj,schumacher,barnum}; if this conjecture is true, then eq.~(\ref{coherent_info}) provides a lower bound on $C_Q$.

Using our continuous variable codes, rigorous lower bounds on $C_Q$ can be established. For $\sigma^2$ sufficiently small, a nonzero attainable rate can be established asymptotically for large $N$ by either of two methods. In one method, the $n=2$ code described in \S\ref{sec:sing_osc} is invoked for each oscillator, and concatenated with a binary quantum code. In the other method, which more closely follows Shannon's construction, a code for $N$ oscillators is constructed as in \S\ref{sec:many_osc}, based on a close packing of spheres in $2N$-dimensional phase space. However (in contrast to the classical case), neither method works if $\sigma^2$ is too large. For large $\sigma^2$, encodings can be chosen that protect against $q$ shifts or against $p$ shifts, but not against both.

To establish an attainable rate using concatenated coding (the method that is easier to explain), we first recall a result concerning the quantum capacities of binary channels \cite{cal_shor,ibm}. If $X$ and $Z$ errors are independent and each occur with probability $p_e$, 
then binary CSS codes exist that achieve a rate
\begin{eqnarray}
R &> &1- 2H_2(p_e)\nonumber\\
&\equiv & 1 + 2p_e\log_2p_e + 2(1-p_e)\log_2 (1-p_e)~;
\end{eqnarray}
this rate is nonzero for $p_e<.1100$.

Now, for the Gaussian quantum channel, if we use the $n=2$ continuous variable code, errors afflicting the encoded qubit are described by a binary channel with independent $X$ and $Z$ errors. Since the code can correct shifts in $q$ or $p$ that satisfy $\Delta q, \Delta p < \sqrt{\pi}/2$, the error probability is
\begin{equation}
\label{gauss_error}
p_e < 2\cdot {1\over \sqrt{2\pi\sigma^2}}\int_{\sqrt{\pi}/2}^\infty dx ~e^{-x^2/2\sigma^2}~.
\end{equation}
Since the expression bounding $p_e$ in eq.~(\ref{gauss_error}) has the value .110 for $\sigma\approx .555$, we conclude that the Gaussian quantum channel has nonvanishing quantum capacity $C_Q$ provided that
\begin{equation}
\sigma < .555~.
\end{equation}

One might expect to do better by concatenating the {\em hexagonal} $n=2$ single-oscillator code with a binary stabilizer code, since the hexagonal code can correct larger shifts than the code derived from a square lattice. For the Gaussian quantum channel, the symmetry of the hexagonal lattice ensures that $X$, $Y$, and $Z$ errors afflicting the encoded qubit are equally likely. 
A shift is correctable if it lies within the ``Voronoi cell'' of the dual lattice, the cell containing all the points that are closer to the origin than to any other lattice site. By integrating the Gaussian distribution over the hexagonal Voronoi cell, we find that the probability $p_{e,{\rm total}}$ of an uncorrectable error satisfies
\begin{equation}
\label{hex_prob}
p_{e,{\rm total}} < 1 - {12\over 2\pi\sigma^2}\int_0^r dx\int_0^{x/\sqrt{3}}dy \, e^{-(x^2+y^2)/2\sigma^2}~,
\end{equation}
where $r=(\pi/2\sqrt{3})^{1/2}$ is the size of the smallest uncorrectable shift. For a binary quantum channel with equally likely $X$, $Y$, and $Z$ errors, it is known \cite{shor_smolin} that there are stabilizer codes achieving a  nonvanishing rate for $p_{e,{\rm total}} < .1905$; our bound on $p_{e,{\rm total}}$ reaches this value for $\sigma\approx .547$. 

Somewhat surprisingly, for very noisy Gaussian quantum channels, square lattice codes concatenated with CSS codes seem to do better than hexagonal codes concatenated with stabilizer codes. The reason this happens is that a CSS code can correct independent $X$ and $Z$ errors that occur with total probability $p_{e,{\rm total}} = p_X + p_Z - p_X\cdot p_Z$, which approaches $.2079 > .1905$ as $p_X=p_Z\to .1100.$ For a given value of $\sigma$, the qubit encoded in each oscillator will have a lower error probability if the hexagonal code is used. But if the square lattice is used, a higher qubit error rate is permissible, and this effect dominates when the channel is very noisy.

We remark that this analysis is readily extended to more general Gaussian quantum channels. We may consider Pauli channels acting on a single oscillator in which the probability distribution $P(\alpha,\beta)$ is a more general Gaussian function, not necessarily symmetric in $p$ and $q$. In that case, a symplectic transformation (one preserving the commutator of $p$ and $q$) can be chosen that transforms the covariance matrix of the Gaussian to a multiple of the identity; therefore, this case reduces to that already discussed above. We may also consider channels acting on $N$ oscillators that apply shifts in the $2N$-dimensional phase space, chosen from a Gaussian ensemble. Again there is a symplectic transformation that diagonalizes the covariance matrix; therefore, this case reduces to $N$ independent single oscillator channels, each with its own value of $\sigma^2$ \cite{referee}. 
 
\section{Symplectic operations}
\label{sec:symplectic}

To use these codes for fault-tolerant quantum computation, we will need to be able to prepare encoded states, perform error recovery, and execute quantum gates that act on the encoded quantum information. The most difficult task is encoding; we will postpone the discussion of encoding until after we have discussed encoded operations and error recovery.

Suppose, for example, that we have $N$ oscillators, each encoding a qunit. We wish to apply $U(n^N)$ transformations that preserve the code subspace of the $N$ qunits. As is typical of quantum codes, we will find that there is a discrete subgroup of $U(n^N)$ that we can implement ``easily;''  but to complete a set of universal gates we must add further transformations that are ``difficult.'' In the case of our continuous variable codes, the easy gates will be accomplished using linear optical elements (phase shifters and beam splitters), along with elements that can ``squeeze'' an oscillator. For the ``difficult'' gates we will require the ability to count photons.

The easy gates are the gates in the Clifford group. In general, the Clifford group of a system of $N$ qunits is the group of unitary transformations that, acting by conjugation, take tensor products of Pauli operators to tensor products of Pauli operators (one says that they preserve the ``Pauli group''). Since for $N$ oscillators the tensor products of Pauli operators have the form eq.~(\ref{general_pauli}), the Clifford group transformations, acting by conjugation, are linear transformations of the $p$'s and $q$'s that preserve the canonical commutation relations. Such transformations are called symplectic transformations. The symplectic group has a subgroup that preserves the photon number
\begin{equation}
{\rm total~ photon~ number}=\sum_{i=1}^N a_i^\dagger a_i~.
\end{equation}
The transformations in this subgroup can be implemented with linear optics \cite{reck}. The full symplectic group also contains ``squeeze operators'' that take an $a$ to a linear combination of $a$'s and $a^\dagger$'s; equivalently, the squeeze operators rescale canonical operators by a real number $\lambda$ along one axis in the quadrature plane, and by $\lambda^{-1}$ along the conjugate axis, as in (for example)
\begin{equation}
q_1\to \lambda q_1~,\quad p_1\to \lambda^{-1} p_1~.
\end{equation}
With squeezing and linear optics we can in principle implement any symplectic transformation.

Aside from the symplectic transformations, we will also assume that it is easy to do displacements that shift $q$ and $p$ by constants. A displacement of $q_1$ by $c$ is actually the limiting case of a symplectic transformation on two oscillators $q_1$ and $q_2$:
\begin{eqnarray}
&q_1 \rightarrow q_1 + \varepsilon q_2~,\quad & p_1\to p_1 + \varepsilon p_2\nonumber \\
&q_2 \rightarrow q_2 - \varepsilon q_1~,\quad & p_2\to p_2 - \varepsilon p_1
\end{eqnarray}
where $\varepsilon \rightarrow 0$ with $\varepsilon q_2 = c$ held fixed. 

Since for the code with stabilizer generators eq.~(\ref{alpha_code}) the Pauli operators acting on our encoded qunits are $\bar X= e^{ip\alpha}$ and $\bar Z=e^{2\pi i q/n\alpha}$, the Clifford group transformations acting on $N$ qunits constitute a subgroup of the symplectic transformations (including shifts) acting on $N$ oscillators, the subgroup that preserves a specified lattice in phase space. Thus we can do any encoded Clifford group gate we please by executing an appropriate symplectic transformation (possibly including a shift).

A similar comment applies to the case of a qunit encoded in a qudit. Since the logical Pauli operators are $\bar X= X^{r_1}$ and $\bar Z= Z^{r_2}$, each Clifford group transformation in the $n$-dimensional code space is also a Clifford group transformation on the underlying qudit. 

But we must also be sure that our implementation of the Clifford group is {\em fault tolerant}. In previous discussions of quantum fault tolerance for $[[N,k,2t+1]]$ codes, the central theme has been that propagation of error from one qudit to another in the same code block must be very carefully controlled \cite{shor_ft,gott_ft}. For shift-resistant codes the main issue is rather different. Since each qudit typically has a (small) error anyway, propagation of error from one qudit to another is not necessarily so serious. But what must be controlled is {\em amplification} of errors -- gates that turn small errors into large errors should be avoided.

The Clifford group can be generated by gates that are fault-tolerant in this sense. 
The Clifford group for qunits can be generated by three elements. The SUM gate is a two-qunit gate that acts by conjugation on the Pauli operators according to 
\begin{equation}
{\rm SUM}: \quad X_1^aX_2^b\to X_1^a X_2^{b-a}~, \quad Z_1^aZ_2^b\to Z_1^{a+b}Z_2^b ~.
\end{equation}
Here qunit 1 is said to be the control of the SUM gate, and qunit 2 is said to be its target; in the binary ($n=2$) case, SUM is known as controlled-NOT, or CNOT. The Fourier gate $F$ acts by conjugation as 
\begin{equation}
F:\quad X\to Z~,\quad Z\to X^{-1}~;
\end{equation}
for $n=2$ the Fourier Transform is called the Hadamard gate. The phase gate $P$ acts as
\begin{equation}
P:\quad X\to (\eta)XZ~,\quad Z\to Z~,
\end{equation}
where the $n$-dependent phase $\eta$ is $\omega^{1/2}$ if $n$ is even and 1 if $n$ is odd. Any element of the Clifford group can be expressed as a product of these three generators. (In Ref.~\cite{higher} another gate $S$ was included among the generators of the Clifford group, but in fact the $S$ gate can be expressed as a product of SUM gates.)

For an $n$-dimensional system encoded in a continuous variable system, these Clifford group generators can all be realized as symplectic transformations. In the case where the stabilizer generators are symmetric in $q$ and $p$, 
\begin{equation}
\bar X= \exp\left (-ip\sqrt{2\pi\over n} \right)~,\quad \bar Z= \exp\left(iq\sqrt{2\pi\over n} \right)~,
\end{equation}
the required symplectic transformations are
\begin{eqnarray}
&{\rm SUM}:\quad & q_1\to q_1~, \qquad\qquad  p_1\to p_1-p_2~,\nonumber\\
& & q_2\to q_1+q_2~, \qquad  p_2\to p_2~,\nonumber\\
&F:\quad & q\to p~,\qquad\qquad \quad p\to -q~,\nonumber\\
&P:\quad & q\to q~,\qquad\qquad \quad p\to p-q+c~,
\end{eqnarray}
where the $n$-dependent shift $c$ is $0$ for $n$ even and $\sqrt{\pi/2n}$ for $n$ odd. Under these symplectic transformations, small deviations of $q$ and $p$ from the stabilizer lattice remain small; in this sense the transformations are fault tolerant. 

\section{Error recovery}
\label{sec:recovery}

If we are willing to destroy the encoded state, then measuring the encoded $\bar X$ or $\bar Z$ is easy -- we simply conduct a homodyne measurement of the $q$ or $p$ quadrature of the oscillator. For example, suppose that we measure $q$ for a state in the code subspace. If there are no errors and the measurement has perfect resolution, the only allowed values of $q$ will be integer multiples of $\alpha$. If there are errors or the measurement is imperfect, classical error correction can be applied to the outcome, by adjusting it to the nearest $\alpha\cdot k$, where $k$ is an integer. Then the outcome of the measurement of $\bar Z$ is $\omega^k$.

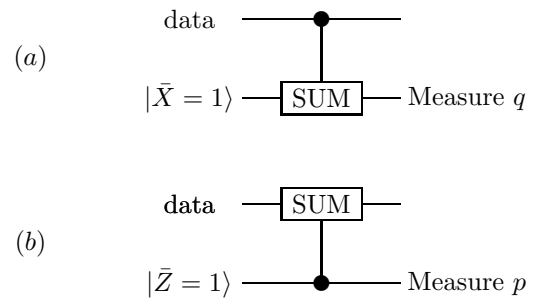
\begin{figure}%[h]
\centering
\begin{picture}(190,120)

%a
\put(0,89){\makebox(20,12){$(a)$}}

\put(60,104){\makebox(20,12){data}}
\put(60,74){\makebox(20,12){$|\bar X=1\rangle$}}

\put(90,80){\line(1,0){15}}
\put(90,110){\line(1,0){60}}

\put(120,110){\circle*{6}}
\put(120,110){\line(0,-1){24}}
\put(105,74){\framebox(30,12){${\rm SUM}$}}
\put(135,80){\line(1,0){15}}

\put(160,74){\makebox(30,12){Measure $q$}}

%b
\put(0,19){\makebox(20,12){$(b)$}}

\put(60,34){\makebox(20,12){data}}
\put(60,34){\makebox(20,12){data}}
\put(60,04){\makebox(20,12){$|\bar Z=1\rangle$}}

\put(90,10){\line(1,0){60}}
\put(90,40){\line(1,0){15}}

\put(120,10){\circle*{6}}
\put(120,10){\line(0,1){24}}
\put(105,34){\framebox(30,12){${\rm SUM}$}}
\put(135,40){\line(1,0){15}}

\put(160,04){\makebox(30,12){Measure $p$}}

\end{picture}
\caption{Measurement of the error syndrome. $(a)$ To diagnose the $q$ shift, an ancilla is prepared in the encoded $\bar X=1$ state, a ${\rm SUM}$ gate is executed with the data as control and the ancilla as target, and the position of the ancilla is measured. $(b)$ To diagnose the $p$ shift, the ancilla is prepared in the $\bar Z=1$ state, a ${\rm SUM}$ gate is executed with the ancilla as control and the data as target, and the momentum of the ancilla is measured.}
\label{fig_syndrome}
\end{figure}

To diagnose errors in a coded data state, we must measure the stabilizer generators. This measurement can be implemented by ``feeding'' the errors from the code block to a coded ancilla, and then measuring the ancilla destructively, following the general procedure proposed by Steane \cite{steane_anc} (see Fig.~\ref{fig_syndrome}). For example, to measure the generator $e^{2\pi i q/\alpha}$ ({\it i.e.}, the value of $q$ modulo $\alpha$), we prepare the ancilla in the state $(|\bar 0\rangle + |\bar 1\rangle)/\sqrt{2}$, the equally weighted superposition of all $|q=s\cdot \alpha\rangle$, $s$ an integer. Then a SUM gate is executed with the data as control and the ancilla as target -- acting according to 
\begin{equation}
q_2\to q_1 + q_2~,
\end{equation}
where $q_1, q_2$ are the values of $q$ for the data and ancilla respectively, prior to the execution of the SUM gate. By measuring $q$ of the ancilla, the value of $q_1 + q_2$ is obtained, and this value modulo $\alpha$ determines the shift that should be applied to the data to recover from the error.

Similarly, to measure the stabilizer generator $e^{inp\alpha}$, we prepare the ancilla in the state $|\bar 0\rangle$, the equally weighted superposition of all $|p=s\cdot 2\pi/n\alpha\rangle$, $s$ an integer. Then a SUM gate is executed with the ancilla as control and the data as target. Finally, the $p$ quadrature of the ancilla is measured. The outcome reveals the value of $p_2-p_1$ prior to the SUM gate, where $p_1$ is the momentum of the data, and $p_2$ is the momentum of the ancilla. The measured value modulo $2\pi/n\alpha$ then determines the shift that should be applied to the data to recover from the error.

Of course, the ancilla used in the syndrome measurement can also be faulty, resulting in errors in the syndrome and imperfect recovery. Similarly, the measurement itself will not have perfect resolution, and the shift applied to recover will not be precisely correct.  Furthermore, as is discussed in \S\ref{sec:finite}, the ideal codewords are unphysical nonnormalizable states, so that the encoded information will always be carried by approximate codewords. For all these reasons, deviations from the code subspace are unavoidable. But if a fresh supply of ancilla oscillators is continuously available, we can prevent these small errors from accumulating and eventually damaging the encoded quantum information.

\section{Universal quantum computation}
\label{sec:universal}
Symplectic transformations together with homodyne measurements are adequate for Clifford group computation and for error recovery (assuming we have a supply of encoded states). But to achieve universal computation in the code space, we need to introduce additional operations. Fortunately, the quantum optics laboratory offers us another tool that can be used to go beyond the symplectic computational model -- the ability to count photons. 

There are a variety of ways in which photon counting can be exploited to complete a universal set of fault-tolerant gates. We will describe two possible ways, just to illustrate how universal fault-tolerant quantum computation might be realized with plausible experimental tools. For this discussion, we will consider the binary case $n=2$.

\subsection{Preparing a Hadamard eigenstate}

We can complete the universal gate set if we have the ability to prepare eigenstates of the Hadamard operator $H$ \cite{knill,ike_dan}. For this purpose it suffices to be able to {\em destructively} measure $H$ of an encoded qubit. Assuming we are able to prepare a supply of the encoded $\bar Z$ eigenstate $|\bar 0\rangle$, we can make an encoded EPR pair using symplectic gates. Then by destructively measuring $H$ for one encoded qubit in the pair, we prepare the other qubit in an encoded eigenstate of $H$ with known eigenvalue.

But how can we destructively measure $H$? The Hadamard gate acts by conjugation on the encoded Pauli operators according to
\begin{equation}
H:\quad \bar X\to \bar Z~, \quad \bar Z\to \bar X~.
\end{equation}
If we use the code that treats $q$ and $p$ symmetrically so that $\bar X= \exp(-ip\sqrt{\pi})$ and $\bar Z= \exp(iq\sqrt{\pi})$, then the Hadamard gate can be implemented by the symplectic transformation.
\begin{equation}
q\to p~,\quad p\to -q~
\end{equation}
(recalling that $\bar X^2=\bar Z^2=I$ on the code subspace). This transformation is just the Fourier transform
\begin{equation}
F:\quad \exp\left( i {\pi\over 2} a^\dagger a\right)
\end{equation}
(where $a^\dagger a$ is the photon number),
which describes the natural evolution of the oscillator for one quarter cycle. Thus the phase of the Hadamard operator is simply the photon number modulo four; we can measure the eigenvalue of the encoded Hadamard transformation by counting photons. 

In fact the photon number in the code space is even -- all codewords are invariant under a $180^\circ$ rotation in the quadrature plane. Because of this feature, the preparation of the Hadamard eigenstate has some fault tolerance built in; if the photon count is off by one, the number will be odd and an error will be detected. In that case we reject the state we have prepared and make a new attempt. If the photon number is large, then obtaining a reliable determination of the photon number modulo four will require highly efficient photodetection. But on the other hand, the photon number need not be very large -- the mean value of $a^\dagger a$ is about $\Delta^{-2}$ where $\Delta$ is the squeeze factor, and we have seen that the intrinsic error rate due to imperfect squeezing is quite small for $\Delta \sim 1/4$, or $\langle a^\dagger a\rangle \sim 16$.

An alternative to preparing an encoded EPR pair and destructively measuring one member of the pair is to prepare $|\bar 0\rangle$ and then perform a quantum nondemolition measurement of the photon number modulo 4. This might be done by coupling the oscillator to a two-level atom as proposed in Ref.~\cite{milburn}. Indeed, since only one bit of information needs to be collected (the photon number is either 0 or 2 modulo 4), the measurement could be made in principle by reading out a single atom. Suppose that the coupling of oscillator to atom is described by the perturbation
\begin{equation}
H'=\lambda ~a^\dagger a ~\sigma_z~,
\end{equation}
where $\sigma_z=-1$ in the atomic ground state $|g\rangle$ and $\sigma_z=1$ in the atomic excited state $|e\rangle$. By turning on this coupling for a time $t=\pi/4\lambda$, we execute the unitary transformation
\begin{equation}
U=\exp( -i(\pi/4)~a^\dagger a~\sigma_z)~.
\end{equation}
Then the atomic state $(|g\rangle + |e\rangle)/\sqrt{2}$ evolves as
\begin{eqnarray}
U:~& &\quad {1\over\sqrt{2}}(|g\rangle + |e\rangle)\nonumber\\
& &\to {1\over\sqrt{2}}e^{ia^\dagger a\pi/4}(|g\rangle + e^{-ia^\dagger a\pi/2}|e\rangle)~.
\end{eqnarray}
By measuring the atomic state in the basis $(|g\rangle \pm |e\rangle)/\sqrt{2}$, we read out the value of the photon number modulo 4 (assumed to be either 2 or 4). Since this is a nondemolition measurement, it can be repeated to improve reliability. By measuring the photon number mod 4 many times (perhaps with rounds of error correction in between the measurements), we obtain a Hadamard eigenstate with excellent fidelity.

How does the ability to construct the Hadamard eigenstate enable us to achieve universal quantum computation? We can make contact with constructions that have been described previously in the literature by observing that the Hadamard eigenstate can be transformed by applying symplectic gates to the ``$\pi/8$ phase state.'' First note that the two Hadamard eigenstates can be converted to one another by applying the encoded gate $\bar X \bar Z$, which can be implemented by shifting both $p$ and $q$. Therefore it is sufficient to consider the eigenstate corresponding to the
eigenvalue $1$,
\begin{equation}
|\psi_{H=1}\rangle\, =\, 
\cos(\pi/8)\,|0\rangle+\sin(\pi/8)\,|1\rangle~.
\end{equation}
By applying the symplectic single-qubit gate
\begin{eqnarray}
H\cdot P^{-1}& \equiv &{1\over\sqrt{2}}\pmatrix{1&1\cr 1& -1\cr}\cdot
\pmatrix{1 & 0\cr 0 & -i\cr}\nonumber\\
& = &{1\over\sqrt{2}}\pmatrix{1 & -i\cr
1 & i\cr}~, 
\end{eqnarray}
we obtain the $\pi/8$ state
\begin{equation}
|\psi_{\pi/8}\rangle\,=\,\frac{1}{\sqrt{2}}\Bigl(e^{-i\pi/8}|0\rangle+e^{i\pi/8}\,|1\rangle\Bigr)~.
\end{equation}

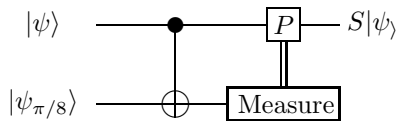
\begin{figure}%[h]
\centering
\begin{picture}(130,55)
\put(0,39){\makebox(20,12){$|\psi\rangle$}}
\put(0,09){\makebox(20,12){$|\psi_{\pi/8}\rangle$}}

\put(30,15){\line(1,0){50}}
\put(30,45){\line(1,0){65}}

\put(60,45){\circle*{6}}
\put(60,45){\line(0,-1){35}}
\put(60,15){\circle{10}}

\put(80,09){\framebox(42,12){\, Measure }}
\put(95,39){\framebox(12,12){$P$}}
\put(107,45){\line(1,0){15}}
\put(125,39){\makebox(20,12){$S|\psi_\rangle$}}

\put(100,39){\line(0,-1){18}}
\put(102,39){\line(0,-1){18}}

\end{picture}
\caption{Implementation of the $S$ gate. An ancilla is prepared in the state $|\psi_{\pi/8}\rangle$, and a CNOT gate is executed with the data as control and the ancilla as target; then the ancilla is measured in the basis $\{|0\rangle$, $|1\rangle\}$. A $P$ gate is applied to the data conditioned on the measurement outcome.}
\label{fig_sgate}
\end{figure}

Now this $\pi/8$ state can be used to perform the nonsymplectic phase gate
\begin{equation}
S=\pmatrix{ e^{-i\pi/8} & 0\cr 0 & e^{i\pi/8}\cr}~,
\end{equation}
which completes the universal gate set \cite{mor,leung}. The gate is constructed by executing the circuit shown in Fig.~\ref{fig_sgate}. We perform a CNOT gate with the arbitrary single-qubit state $|\psi\rangle=a|0\rangle + b|1\rangle$ as the control, and the $\pi/8$ phase state as the target; then the target qubit is measured in the basis $\{|0\rangle,|1\rangle\}$. If the measurement outcome is $|0\rangle$ (which occurs with probability 1/2), then the control qubit has become $ae^{i\pi/8}|0\rangle + b e^{-i\pi/8} |1\rangle= S|\psi\rangle$ and we are done. If the measurement outcome is $|1\rangle$, then the control qubit has become  $ae^{-i\pi/8}|0\rangle + b e^{i\pi/8} |1\rangle$, and we obtain $S|\psi\rangle$ by applying the symplectic single-qubit gate
\begin{equation}
e^{-i\pi/4} P = \pmatrix {e^{-i\pi/4} & 0\cr 0 & e^{i\pi/4}\cr}~.
\end{equation}

Completing the universal gate set by measuring the Hadamard transformation has some drawbacks. For one thing, while photon number modulo four corresponds to the Hadamard eigenvalue in the ideal code space, this correspondence will not apply to approximate codewords unless they are of a special type. 

Recall that the imperfections of the codewords arising from finite squeezing can be described by an ``embedded error'' $|\eta\rangle$ as in eq.~(\ref{error_wave_function}); a Gaussian approximate codeword has a Gaussian embedded error 
\begin{equation}
%\label{Gaussian_error}
\eta(u,v)\, =\, \frac{1}{\sqrt{\pi\Delta\kappa}}
\exp\Bigl(-\frac{1}{2}(u^2/\Delta^2+v^2/\kappa^2)\Bigr),
\end{equation}
where $\Delta$ is the width in $q$ and $\kappa$ is the width in $p$.
Symplectic gates act separately on the encoded qubit and the ``embedded
error'' $|\eta\rangle$; for example, the Fourier transform gate and the ${\rm
SUM}$ gate act on the error according to
\begin{eqnarray}
F: & |u,v\rangle\, &\to\, |v,-u\rangle ~,\nonumber\\
{\rm SUM}: & |u_1,v_1;u_2,v_2\rangle\, &\to\,
|u_1,\,v_1+v_2;\,u_2-u_1,\,v_2\rangle~.
\end{eqnarray}
By measuring the photon number modulo $4$, we actually measure the {\em product}
of the eigenvalue of the Hadamard gate acting on the codeword and the eigenvalue of $F$ acting on the embedded error. The latter always equals $1$ {\em if} we use symmetrically squeezed
codewords, with  $\Delta=\kappa$.

Symmetric squeezing is not in itself sufficient to ensure that the measurement of the photon number modulo 4 will prepare the desired encoded Hadamard eigenstate. We also need to consider how the embedded error is affected by the preparation of the EPR pair that precedes the measurement. To prepare the EPR pair, we use the SUM gate. Suppose that we start with
two symmetrically squeezed states. Then the SUM gate yields the
error wave function
\begin{eqnarray}
& & \eta'(u_1,v_1;u_2,v_2)\, \nonumber\\
& & =\,\exp \left(- \left(u_1^2+(v_1-v_2)^2+(u_1+u_2)^2+v_2^2\right) /\Delta^2\right)~.
\end{eqnarray}
Not only is it not symmetric, but the error is entangled between the two
oscillators. The Fourier transform measurement will not give the desired
result when applied to either oscillator.

To ameliorate this problem, we could perform error correction after the preparation of the EPR pair and before the measurement, where the error correction protocol has been designed to produce symmetrically squeezed states. Or we could avoid preparing the EPR state by using the nondemolition measurement of photon number modulo 4, as described above.

\subsection{Preparing a cubic phase state}

Now we will describe another way to use photon counting to implement non-symplectic gates, which is less sensitive to the codeword quality. Again, we will complete the universal gate set by constructing the $\pi/8$ phase gate $S$.

For our binary ($n=2$) code, the code subspace has the basis
\begin{eqnarray}\label{binary_codewords}
|\bar 0\rangle\,& = &\, \sum_{s=-\infty}^{+\infty} |q=2s\alpha\rangle~,\nonumber\\
|\bar 1\rangle\, &= &\, \sum_{s=-\infty}^{+\infty} |q=(2s+1)\alpha\rangle~.
\end{eqnarray}
(For now we ignore the  embedded error due to imperfect squeezing; it will be taken into account later.) An $S$ gate acting on the encoded qubit is implemented (up to an irrelevant overall phase) by the unitary operator
\begin{equation}\label{W}
W\,=\, \exp\left({i\pi\over 4}\left[ 
2(q/\alpha)^3+(q/\alpha)^2-2(q/\alpha)\right]\right)~.
\end{equation}
Indeed, we can check that 
\begin{equation}
2x^3+ x^2-2x\ (\bmod\ 8)\, =\, \cases{0,& if $x=2s$~,\cr 1,& if $x=2s+1$~.\cr}
\end{equation}

\begin{figure}%[h]
\centering
\begin{picture}(210,75)
\put(0,54){\makebox(20,12){$|\psi\rangle$}}
\put(0,14){\makebox(20,12){$|\gamma\rangle$}}

\put(30,20){\line(1,0){20}}
\put(30,60){\line(1,0){100}}
\put(160,60){\line(1,0){20}}
\put(90,20){\line(1,0){25}}

\put(70,60){\circle*{6}}
\put(70,60){\line(0,-1){30}}

\put(50,10){\framebox(40,20){${\rm SUM}^{-1}$}}
\put(130,50){\framebox(30,20){$U(a)$}}
\put(115,10){\framebox(60,20){Measure $q$}}

\put(190,54){\makebox(20,12){$V_\gamma|\psi_\rangle$}}

\put(144,50){\line(0,-1){20}}
\put(146,50){\line(0,-1){20}}

\end{picture}
\caption{Implementation of the cubic phase gate. An ancilla is prepared in the state $|\gamma\rangle$, and a ${\rm SUM}^{-1}$ gate is executed with the data as control and the ancilla as target; then the position of the ancilla is measured. A symplectic gate $U(a)$ is then applied to the data, conditioned on the outcome $a$ of the measurement.}
\label{fig_phigate}
\end{figure}
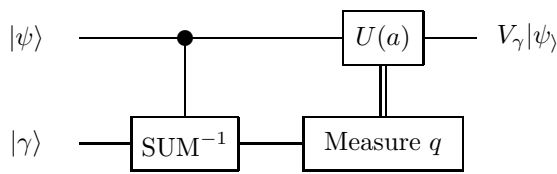

The operator $W$ is the product of a symplectic gate and the cubic phase gate
\begin{equation}\label{V}
V_\gamma\, =\, \exp(i\gamma q^3)~,
\end{equation}
where $\gamma=\pi/(2\alpha^3)$. But how do we implement the cubic gate? In fact, if we are able to prepare a ``cubic phase state''
\begin{equation}
|\gamma\rangle = \int dx \,  e^{i\gamma x^3}|x\rangle~,
\end{equation}
then we can perform the gate $V_\gamma$ by executing the circuit shown in Fig.~\ref{fig_phigate}.

To understand how the circuit works, consider the more general problem of implementing a phase gate that acts on the position eigenstates according to
\begin{equation}
V_\phi: |q\rangle \to e^{i\phi(q)}|q\rangle
\end{equation}
(where $\phi(q)$ is a real-valued function), using the prepared phase state
\begin{equation}
|\phi\rangle = \int dx \, e^{i\phi(x)}|x\rangle~.
\end{equation}
If we perform the gate ${\rm SUM}^{-1}$ with position eigenstate $|q\rangle$ as control and $|\phi\rangle$ as target, and then measure the position of the target obtaining the outcome $|a\rangle$, the state of the control oscillator has become $e^{i\phi(q+a)}|q\rangle$. We can therefore complete the construction of $V_\phi$ by applying the transformation
\begin{equation}
U(a) = e^{i[\phi(q)-\phi(q+a)]}~.
\end{equation}
If the function $\phi(q)$ is cubic, then the argument of the exponential is quadratic and hence $U(a)$ is a symplectic transformation.

Now the problem of implementing universal quantum computation in the code subspace has been reduced to the problem of preparing the cubic phase state $|\gamma\rangle$. We can accomplish this task by preparing an EPR pair, and then performing a suitable photon counting measurement (a nonideal homodyne measurement) on one member of the pair.

Of course, the EPR pair will not be perfect. To be definite, let us suppose (although this assumption is not really necessary) that it is a Gaussian state
\begin{eqnarray}
\label{psi_kappa_delta}
|\psi_{\sigma_p,\sigma_q}\rangle & = & \left({\sigma_p\over \pi\sigma_q}\right)^{1/2}\int dq_1 dq_2  ~\exp \left[- {1\over 2}\sigma_p^2\left({q_1+q_2\over 2}\right)^2\right] \nonumber\\
& &\times \exp\left[ - {1\over 2}\left(q_1-q_2\right)^2/\sigma_q^2\right]|q_1,q_2\rangle 
\end{eqnarray}
with $\sigma_p,\sigma_q \ll 1$. 

Now suppose that the second oscillator is mixed with a coherent light beam, resulting in a large shift in momentum,
\begin{equation}
|\psi\rangle \to e^{iwq}|\psi\rangle~,\quad w\gg \sigma_q^{-1},\sigma_p^{-1}~;
\end{equation}
then the photon number is measured and $n$ photons are detected. Thus the state of the first oscillator becomes (up to normalization)
\begin{eqnarray}
\label{psi_one}
|\psi_{1}^{(n)}\rangle & \approx & \left({\sigma_p\over \pi\sigma_q}\right)^{1/2}\int dq_1\, |q_1\rangle \,  
e^{-{1\over 2}\sigma_p^2 q_1^2}\, \nonumber\\
& &\times \int dq_2\, \varphi_n^*(q_2) e^{iwq_2} 
e^{-{1\over 2}\left(q_1-q_2\right)^2/\sigma_q^2}~,
\end{eqnarray}
where $|\varphi_n\rangle$ denotes the photon number eigenstate, the eigenstate with eigenvalue $n+{1\over 2}$ of the Hamiltonian $H={1\over 2}(p^2+q^2)$. 

We can evaluate the $q_2$ integral in eq.~(\ref{psi_one}) by appealing to the semiclassical approximation. For $q_2$ in the classically allowed region and far from the classical turning points, we may write
\begin{eqnarray}
\varphi_n^*(q_2) &\sim &{1\over \sqrt{2\pi p(q_2)}}\exp\left(-i \int^{q_2} dx\, p(x)\right)\nonumber\\
&&+{1\over \sqrt{2\pi p(q_2)}}\exp\left(+i \int^{q_2} dx\, p(x)\right)~,
\end{eqnarray}
where
\begin{equation}
p(x)= \sqrt{2n+1 - x^2}~.
\end{equation}
For $w\gg\sigma_q^{-1}$, the rapid phase oscillations strongly suppress the contribution to the integral arising from the left-moving part of $\varphi^{(n)}(q_2)$. A contribution from the right-moving part survives provided that 
\begin{equation}
\label{q1_condition}
|p(q_1)-w| < \sigma_q^{-1}~.
\end{equation}
When this condition is satisfied, it is a reasonable approximation to replace the Gaussian factor $e^{-{1\over 2}\left(q_1-q_2\right)^2/\sigma_q^2}$ in the $q_2$ integral by $\sqrt{2\pi\sigma_q^2}~\delta(q_1-q_2)$, so that we obtain
\begin{eqnarray}
|\psi_{1}^{(n)}\rangle & \approx & \left(2\sigma_p\sigma_q\right)^{1/2}\int dq_1\, |q_1\rangle \,  
e^{-{1\over 2}\sigma_p^2 q_1^2}\, \nonumber\\
 \times &&{1\over \sqrt{2\pi p(q_1)}}\exp\left(-i \int^{q_1} dx\, \left(p(x)-w\right)\right)~.
\end{eqnarray}
The probability that $n$ photons are detected is given by the norm of this $|\psi_1^{(n)}\rangle$. The values of $n$ that occur with appreciable probability satisfy eq.~(\ref{q1_condition}) for some $q_1$ with $|q_1| < \sigma_p^{-1}$; thus typical measurement outcomes are in the range 
\begin{equation}
n+{1\over 2}\sim {1\over 2}\left(w \pm \sigma_q^{-1}\right)^2 + {1\over 2}\sigma_p^{-2}~,
\end{equation}
with a flat probability distribution
\begin{equation}
{\rm Prob}(n)= \langle \psi_1^{(n)}|\psi_1^{(n)}\rangle \sim {\sigma_q\over w}~.
\end{equation}
Heuristically, after the momentum shift is applied, the oscillator that is measured has momentum of order $w \pm \sigma_q^{-1}$, and position of order $\sigma_p^{-1}$, so that the value of the energy is $n+{1\over 2}={1\over 2}(p^2 + q^2)\sim {1\over 2}\left(w \pm \sigma_q^{-1}\right)^2 + {1\over 2}\sigma_p^{-2}$.

For a particular typical outcome of the photon-counting measurement, since $|\psi_1^{(n)}\rangle$ has its support on $|q_1| <  \sigma_p^{-1} \ll w$, we can Taylor expand $p(x)$ about $x=q_1$ to express $|\psi_1^{(n)}\rangle$ as
\begin{eqnarray}
\label{psi_one_expand}
&&\psi_1^{(n)}(q_1)\propto 
\exp\left(-i\int^{q_1}\bigl(\sqrt{(2n+1)-x^2}-w\bigr)\,dx\right)\nonumber\\
&&\propto
\exp\Biggl(\frac{i}{6\sqrt{2n+1}}q_1^3-i\bigl(\sqrt{2n+1}-w\bigr)q_1\nonumber\\
&&\qquad\qquad\qquad + O(q_1^5/w^3)\Biggr)~.
\end{eqnarray}
This is a cubic phase state to good precision if $w$ is large enough. 

The coefficient $\gamma'$ of $q_1^3$ in the phase of $\psi_1$ is of order $n^{-1/2}$, while the phase $\gamma$  of the operator $V_\gamma$ that we wish to execute is of order one. However, we can construct $V_\gamma$ from $V_{\gamma'}$ as
\begin{equation}
V_\gamma = \left(S_{\gamma/\gamma'}\right)^{-1} V_{\gamma'} \left(S_{\gamma/\gamma'}\right) ~,
\end{equation}
where $S_{r}$ is a squeeze operation that acts according to 
\begin{eqnarray}
\label{gate_squeeze}
S_{r}: q &\to& (r)^{1/3}q~,\nonumber\\
p &\to& (r)^{-1/3}p~.
\end{eqnarray}
Alternatively, we could squeeze the phase state $|\gamma'\rangle$ before we use it to implement the cubic phase gate. 

Is this procedure fault tolerant?  Before considering the errors introduced during the implementation of the cubic phase gate, we should check that the gate does not catastrophically amplify any preexisting errors. In general, a phase gate can transform a small position shift error into a potentially dangerous momentum shift error. Commuting $V(\phi)=e^{i\phi(q)}$ through the shift operator $e^{-iup}$, we find
\begin{equation}
e^{i\phi(q)}e^{-iup}= e^{-iup}e^{i f_u(q)}e^{i\phi(q)}~,
\end{equation}
where $f_u(q)= \phi(q+u)-\phi(q)$; the operator $e^{if_u(q)}$ can be expanded in terms of momentum shift operators of the form $e^{ivq}$ by evaluating the Fourier transform
\begin{equation}
\tilde f_u(v)=\int {dq\over 2\pi} e^{i(f_u(q)-vq)}~.
\end{equation}
Assuming we use a code where the parameter $\alpha$ is of order one, uncorrectable errors will be likely if $\tilde f_u(v)$ has significant support on values of $v$ that are order one.

Suppose that $V(\phi)$ acts on an approximate codeword whose wave function is concentrated on values of $q$ in the domain $|q|< L$. Phase cancellations will strongly suppress $\tilde f_u(v)$, unless the 
stationary phase condition $f_u'(q)=v$ is satisfied for some value of $q$ in the domain of the approximate codeword. Therefore, $V(\phi)$ can propagate a preexisting position shift $u$ to a momentum shift error of magnitude
\begin{equation}
|v|\sim \max_{|q|\le L} |f_u'(q)|~.
\end{equation}
The cubic phase gate needed to implement the encoded $S$ gate is $W=e^{i \phi(q)}$ where $\phi(q)=\pi q^3/2\alpha^3$, so that $f_u(q)= 3\pi u q^2/2\alpha^3 +\cdots$ (ignoring small terms linear and constant in $q$), and  $f_u'(q)=3\pi u q/\alpha^3$; the gate transforms the position shift $u$ to a momentum shift
\begin{equation}
v\sim 3\pi L u/\alpha^3~.
\end{equation}
For $\alpha$ of order one, then, to ensure that $v$ is small we should use approximate codewords with the property that the typical embedded position shift $u$ satisfies
\begin{equation}
\label{u_condition}
|u|\ll L^{-1}~.
\end{equation}
In particular, if the approximate codeword's embedded errors are Gaussian, where $\kappa$ is the typical size of a momentum shift and $\Delta$ is the typical size of a position shift, we require
\begin{equation}
\Delta \ll \kappa~.
\end{equation}
We assume that shift errors due to other causes are no larger than the embedded error.

In the circuit Fig.~\ref{fig_phigate} that implements the cubic phase gate, position shift errors in either the encoded state $|\psi\rangle$ or the ancilla state $|\gamma\rangle$ might cause trouble. A shift by $u$ in $|\psi\rangle$ is transformed to a phase error $e^{if_u(q)}$, and a shift by $u$ in $|\gamma\rangle$ infects $|\psi\rangle$ with a phase error $e^{if_u(q +a)}$. Therefore, we should require that position shift errors in both $|\psi\rangle$ and $|\gamma\rangle$ satisfy the criterion eq.~(\ref{u_condition}), where $L$ is the larger of the two wave packet widths. 

When a cubic phase state is prepared by measuring half of an EPR pair, the packet width is of order $\sigma_p^{-1}$ and typical position shift errors have $u\sim \sigma_q$. However, we must also take into account that either the encoded state or the ancilla must be squeezed as in eq.~(\ref{gate_squeeze}). Suppose that the ancilla is squeezed, by a factor of order $n^{1/6}\sim w^{1/3}$; the wave packet is rescaled so that, after squeezing, the width $L'$ and the typical shifts $u'$ are given by
\begin{equation} 
L'\sim \sigma_p^{-1}w^{-1/3}~, \qquad u'\sim \sigma_q w^{-1/3}~.
\end{equation}
Then the condition $|u'| \ll L'^{-1}$ is satisfied provided that $\sigma_q \ll \sigma_p w^{2/3}$. We also require that the rescaled packet has width large compared to 1, or $\sigma_p \ll w^{-1/3}$.

For the derivation of eq.~(\ref{psi_one_expand}), we used the approximations $w\sigma_q \gg 1$ and $w\sigma_p \gg1$. We also need to check that the remainder terms in the Taylor expansion give rise to a phase error that is acceptably small. This error has the form $e^{if(q_1)}$, where $f(q_1)= O(q_1^5/w^3)$, corresponding to a momentum shift
\begin{equation}
v\sim f'(q_1) \sim \sigma_p^{-4} w^{-3}~.
\end{equation}
Squeezing amplifies this momentum shift error to $v'\sim vw^{1/3}\sim \sigma_p^{-4}w^{-8/3}$, which will be small compared to 1 provided that $\sigma_p \gg w^{-2/3}$. To summarize, our implementation of the cubic phase gate works well if the approximate codewords have embedded errors satisfying $\Delta \ll \kappa$, and if widths $\sigma_q$ and $\sigma_p$ of the approximate EPR state satisfy $w\gg \sigma_q^{-1}$  and 
\begin{equation}
w^{-1/3} \gg \sigma_p \gg w^{-2/3}~.
\end{equation}

Finally, how accurately must we count the photons? An error $\Delta n$ in the photon number results in a phase error $e^{ivq_1}$ with $|v|\sim n^{-1/2}\Delta n$ in $\psi_1^{(n)}(q_1)$, which will be amplified by squeezing to $|v'|\sim |v|w^{1/3}\sim n^{-1/3}\Delta n$. Therefore, the precision of the photon number measurement should satisfy
\begin{equation}
\Delta n \ll n^{1/3}
\end{equation}
to ensure that this error is acceptably small.

\subsection{Purification}

Either of the above two methods could be used to implement a nonsymplectic phase transformation that completes the universal gate set.  Of course, experimental limitations might make it challenging to execute the gate with very high fidelity. One wonders whether it is possible to refine the method to implement fault-tolerant universal gates of improved fidelity.

In fact, such refinements are possible. We have seen that we can reach beyond the symplectic transformations and achieve universal quantum computation if we have a supply of appropriate ``nonsymplectic states'' that can't be created with the symplectic gates. If the nonsymplectic states have the right properties, then we can carry out a purification protocol to distill from our initial supply of noisy nonsymplectic states a smaller number of nonsymplectic states with much better fidelity \cite{kit_purify,dennis}. 

An example of a nonsymplectic state that admits such a purification protocol is a variant of the state originally introduced by Shor \cite{shor_ft}, the three-qubit state
\begin{equation}
2^{-3/2}\sum_{a,b,c\in\{0,1\}} (-1)^{abc}|a\rangle_1|b\rangle_2|c\rangle_3~;
\end{equation}
it can be characterized as the simultaneous eigenstate of three commuting symplectic operators: 
$\Lambda(Z)_{1,2}X_3$ and its two cyclic permutations, where $\Lambda(Z)$ is the two-qubit conditional phase gate
\begin{equation}
\Lambda(Z): |a,b\rangle \to (-1)^{ab}|a,b\rangle
\end{equation}
As Shor explained, this nonsymplectic state can be employed to implement the Toffoli gate
\begin{equation}
T:|a,b,c\rangle \to |a,b,c\oplus ab\rangle~,
\end{equation}
and so provides an alternative way to complete the universal gate set.

To purify our supply of nonsymplectic states, symplectic gates are applied to a pair of nonsymplectic states and then one of the states is measured. Based on the outcome of the measurement, the other state is either kept or discarded. If the initial ensemble of states approximates the nonsymplectic states with adequate fidelity, then as purification proceeds, the fidelity of the remaining ensemble converges rapidly toward one.

\begin{figure}%[h]
\centering
\begin{picture}(260,120)

%a
\put(0,84){\makebox(20,12){$(a)$}}

\put(30,100){\line(1,0){40}}
\put(30,80){\line(1,0){12}}
\put(58,80){\line(1,0){12}}

\put(50,100){\circle*{6}}
\put(50,100){\line(0,-1){12}}
\put(42,72){\framebox(16,16){\small $P$}}

\put(85,84){\makebox(10,12){$=$}}

\put(110,100){\line(1,0){72}}
\put(110,80){\line(1,0){32}}
\put(158,80){\line(1,0){24}}
\put(198,100){\line(1,0){12}}
\put(198,80){\line(1,0){12}}

\put(130,100){\circle*{6}}
\put(130,100){\line(0,-1){25}}
\put(130,80){\circle{10}}

\put(170,100){\circle*{6}}
\put(170,100){\line(0,-1){25}}
\put(170,80){\circle{10}}

\put(142,72){\framebox(16,16){\small $S^{-1}$}}
\put(182,72){\framebox(16,16){\small $S$}}
\put(182,92){\framebox(16,16){\small $S$}}
\put(198,10){\line(1,0){32}}

%b
\put(0,24){\makebox(20,12){$(b)$}}

\put(30,50){\line(1,0){40}}
\put(30,30){\line(1,0){40}}
\put(30,10){\line(1,0){12}}
\put(58,10){\line(1,0){12}}

\put(50,50){\circle*{6}}
\put(50,30){\circle*{6}}
\put(50,50){\line(0,-1){32}}
\put(42,2){\framebox(16,16){\small $Z$}}

\put(85,24){\makebox(10,12){$=$}}

\put(110,50){\line(1,0){120}}
\put(110,30){\line(1,0){92}}
\put(110,10){\line(1,0){32}}

\put(130,30){\circle*{6}}
\put(130,30){\line(0,-1){25}}
\put(130,10){\circle{10}}

\put(142,2){\framebox(16,16){\small $P^{-1}$}}
\put(158,10){\line(1,0){24}}

\put(170,30){\circle*{6}}
\put(170,30){\line(0,-1){25}}
\put(170,10){\circle{10}}

\put(182,2){\framebox(16,16){\small $P$}}
\put(198,10){\line(1,0){32}}

\put(202,22){\framebox(16,16){\small $P$}}
\put(218,30){\line(1,0){12}}

\put(150,50){\circle*{6}}
\put(190,50){\circle*{6}}
\put(210,50){\circle*{6}}

\put(210,50){\line(0,-1){12}}
\put(190,50){\line(0,-1){18}}
\put(150,50){\line(0,-1){18}}

\put(190,28){\line(0,-1){10}}
\put(150,28){\line(0,-1){10}}

\end{picture}
\vskip .1in
\caption{Construction of the three-qubit gate $\Lambda^2(Z)$. $(a)$ A $\Lambda(P)$ gate can be constructed (up to an overall phase) from two $S$ gates, an $S^{-1}$ gate, and two CNOT's. The circuit is executed from left to right. $(b)$ A $\Lambda^2(Z)$ gates can be constructed from two $\Lambda(P)$ gates, a $\Lambda(P^{-1})$ gate, and two CNOT's.}
\label{fig_shor}
\end{figure}
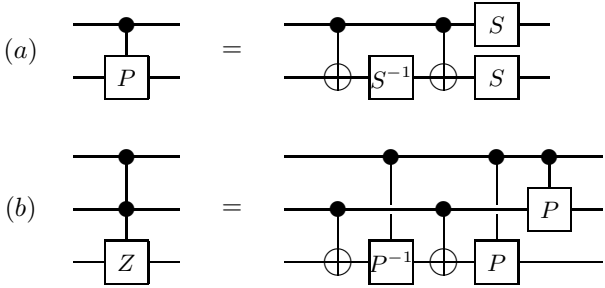

The details of the purification protocol will be described elsewhere; here we will only remark that these Shor states can be readily created using symplectic gates and $\pi/8$ phase gates. The Shor state is obtained if we apply the transformation
\begin{equation}
\Lambda^2(Z): |a,b,c\rangle \to (-1)^{abc} |a,b,c\rangle 
\end{equation}
to the state
\begin{equation}
H_1H_2H_3 |0,0,0\rangle = 2^{-3/2}\sum_{a,b,c\in\{0,1\}} |a,b,c\rangle~.
\end{equation}
As shown in Fig.~\ref{fig_shor}, $\Lambda^2(Z)$ can be applied by executing a circuit containing 5 $S$ gates, 4 $S^{-1}$ gates, and 8 CNOT gates. 

Therefore, if we can apply symplectic gates accurately, and are also able to create a supply of $\pi/8$ states of reasonable fidelity (or can otherwise implement $S$ gates of reasonable fidelity), then we can use the purification protocol to implement Toffoli gates with very good fidelity.

\section{Encoding}
\label{sec:encoding}

Now we have discussed how to execute universal quantum computation fault tolerantly, and how to perform error recovery. But the discussion has all been premised on the assumption that we can prepare encoded states. It is finally time to consider how this can be done. In fact, preparing simultaneous eigenstates of the stabilizer generators $\exp(2\pi iq/\alpha)$ and $\exp(-inp\alpha)$ is a challenging task.  

For the $[[N,k]]$ stabilizer codes that have been discussed previously, encoding is not intrinsically difficult in that it can be accomplished with Clifford group gates. Acting by conjugation, Clifford group transformations take tensor products of Pauli matrices to tensor products of Pauli operators. In particular, there is a Clifford group transformation that takes the state $|0\rangle^{\otimes N}$ (the simultaneous eigenstate with eigenvalue one of all $N$ single-qubit $Z$'s) to the encoded $|\bar 0\rangle^{\otimes k}$ (the simultaneous eigenstate with eigenvalue one of $(N-k)$ stabilizer generators and $k$ encoded $\bar Z$'s). 

Where our codes are different, in both their finite-dimensional and infinite-dimensional incarnations, is that a {\em single} qudit or oscillator is required to obey {\em two} independent stabilizer conditions -- i.e., to be the simultaneous eigenstate of two independent Pauli operators. Hence there is no Clifford group encoder. In the continuous variable case, the problem can be stated in more familiar language: the symplectic transformations take Gaussian (coherent or squeezed) states to Gaussian states. Hence no symplectic transformation can take (say) the oscillator's ground state to a state in the code subspace.

So encoding requires nonsymplectic operations, and as far as we know it cannot be accomplished by counting photons either -- we must resort to a nonlinear coupling between oscillators, such as a $\chi^{(3)}$ coupling. We will describe one possible encoding scheme: First, we prepare a squeezed state, an eigenstate of the momentum with $p=0$. This state is already an eigenstate with eigenvalue one of the stabilizer generator $e^{inp\alpha}$, but not an eigenstate of $e^{2\pi iq/\alpha}$; rather its value of $q$ is completely indefinite. To obtain an encoded state, we must project out the component with a definite value of $q$ modulo $\alpha$.

\begin{figure}
\begin{center}
\leavevmode
\epsfxsize=3in
\epsfbox{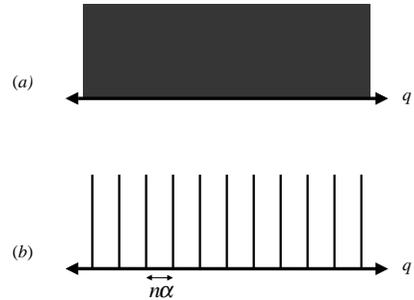}
\end{center}
\caption{Preparation of an encoded state. $(a)$ An eigenstate of $p$ is prepared, which has an indefinite value of $q$. $(b)$ The value of $q$ modulo $n\alpha$ is measured, projecting out a state that differs from the encoded $\bar Z$ eigenstate by a shift in $q$.}
\label{fig_encoding}
\end{figure}

This can be achieved by coupling the oscillator to another oscillator that serves as a meter, via the perturbation of the Hamiltonian
\begin{equation}
H'=\lambda ~ q ~ \left(b^\dagger b\right)~,
\end{equation}
where $b$ is the annihilation operator of the meter.\footnote{There is an extensive literature on the experimental realization and applications of this kind of coupling; see \cite{nonlinear}.} This perturbation modifies the frequency of the meter,
\begin{equation}
\Delta\omega_{\rm meter}= \lambda ~q~;
\end{equation}
then if this coupling is turned on for a time $t=2\pi/\lambda n \alpha$, the phase of the meter advances by 
\begin{equation}
\Delta\theta_{\rm meter}= 2\pi q/ n \alpha~.
\end{equation}
By reading out the phase, we can determine the value of  $q$ modulo $n \alpha$, and apply a shift if necessary to obtain the state with $q\equiv 0$ (mod $n \alpha$), the known state $|\bar 0\rangle$ in the code subspace. (See Fig. \ref{fig_encoding}.)

Of course, in practice the state squeezed in $p$ prepared in the first step will be only finitely squeezed, and the measurement of $q$ modulo $n\alpha$ will have imperfect resolution. If the squeezed state is Gaussian and the measurement has a Gaussian acceptance, then this procedure will produce an approximate codeword of the sort described in \S\ref{sec:finite}.

If we are able to prepare ``good enough'' encoded states, we can distill better ones. The distillation protocol is similar to the error recovery procedure, but where the ancilla used for syndrome measurement may be fairly noisy. We might improve the convergence of the distillation procedure by discarding the data oscillator if the measurement of the ancilla oscillator yields a value of $q$ or $p$ that is too distant from the values allowed by the code stabilizer.

So far, we have described how to prepare encoded states for the ``single-oscillator'' codes described in \S\ref{sec:sing_osc}. To prepare an encoded state for one of the $N$-oscillator codes described in \S\ref{sec:many_osc}, we proceed in two steps. First we prepare each of $N$ oscillators in a single-oscillator encoded state. Then we apply a symplectic transformation to obtain the encoded state of the $N$-oscillator code. 

A particular known encoded state of a lattice stabilizer code can itself be regarded as a code with an $(n=1)$-dimensional code space. Hence it can be characterized by a {\em self-dual} symplectic lattice. For example, the $\bar X=1$ state of a qunit encoded in a single oscillator is the simultaneous eigenstate with eigenvalue one of the operators $e^{-ip\alpha}$ and $e^{2\pi i q/\alpha}$ -- the state associated with the self-dual lattice whose basis vectors are $p\alpha /\sqrt{2\pi}$ and $q\sqrt{2\pi}/\alpha$. 

One encoded state can be transformed to another by symplectic gates if there is a symplectic linear transformation that takes the self-dual lattice associated with the first state to the self-dual lattice associated with the second. In fact, such a symplectic transformation exists for any pair of self-dual lattices.

A linear transformation acting on the $p$'s and $q$'s  modifies the generator matrix $M$ of a lattice according to 
\begin{equation}
M \to MS~;
\end{equation}
this transformation is symplectic if 
\begin{equation}
S\omega S^T = \omega~,
\end{equation}
where 
\begin{equation}
\omega=\pmatrix{0 & I \cr -I & 0}~.
\end{equation}
We saw in \S\ref{sec:many_osc} that we can always choose the generator matrix $M$ of a self-dual lattice so that the matrix $A$ has the form
\begin{equation}
A\equiv  M\omega M^T=\omega~;
\end{equation}
that is, so that $M$ is a symplectic matrix. Therefore, the generator matrices $M_1$ and $M_2$ of two self-dual lattices can each be chosen to be symplectic; then the linear transformation
\begin{equation}
S= M_1^{-1}M_2
\end{equation}
that takes one lattice to the other is also symplectic.

Thus, while the task of preparing the encoded states of the single-oscillator codes can be accomplished only by introducing a nonlinear coupling between oscillators, proceeding from single-oscillator encoded states to many-oscillator encoded states can be achieved with linear optical operations and squeezing.

\section{Physical fault tolerance?}
\label{sec:physical}

In a physical setting, making use of the continuous variable quantum error-correcting codes proposed here (or ``digital'' quantum codes that have been proposed previously) is a daunting challenge. We must continually measure the stabilizer operators (the ``error syndrome'') to diagnose the errors; to recover we must apply frequent shifts of the canonical variables that are conditioned on the measurement outcomes. Cold ancilla oscillators must be provided that are steadily consumed by  the syndrome measurements. The ancillas must be discarded (or refreshed) to rid the system of excess entropy that has been introduced by the accumulated errors.

An alternative to this complex scheme was suggested in Ref.~\cite{kitaev}. Perhaps we can engineer a quantum system whose (degenerate) ground state is the code subspace. Then the natural coupling of the system to its environment will allow the system to relax to the code space, removing errors introduced by quantum and thermal noise,  or through the imperfect execution of quantum gates. Such a system, if it could be built, would be a highly stable quantum memory.

Continuous variable coding suggests new approaches to implementing this type of physical fault tolerance. For example, the Hamiltonian 
\begin{equation}
H=2-\left[\cos p + \cos (2\pi n q)\right]
\end{equation}
has an $n$-fold degenerate (but nonnormalizable) ground state that is just the code space of a continuous variable code. (The operators $\cos p$ and $\cos 2\pi n q$ commute and can be simultaneously diagonalized.) The low-lying states of a real system whose Hamiltonian is a reasonable approximation to $H$ would resemble the approximate codewords described in \S\ref{sec:finite}. 

One possible way to realize physical fault tolerance is suggested by the codes for an electron in a Landau level, described in \S\ref{sec:landau}. The wave functions in the code space are doubly periodic with a unit cell that encloses $n$ flux quanta, where $n$ is the code's dimension. If we turn on a tunable periodic potential whose unit cell matches that of the code, then the Landau level is split into $n$ energy bands, and the codewords are the states with vanishing Bloch momentum. Therefore, an encoded state could be prepared by turning on the potential, waiting for dissipative effects to cause the electrons to relax to the bottom of the lowest band, and then adiabatically turning off the potential. If dissipative effects cause electrons to relax to the bottom of a band on a time scale that is short compared to spontaneous decay from one band to another, then more general encoded states could be prepared by a similar method. Furthermore, turning on the potential from time to time would remove the accumulated Bloch momentum introduced by errors, allowing the electron to relax back to the code space.

\section{Concluding comments}
\label{sec:conclude}

We have described codes that protect quantum states encoded in a finite-dimensional subspace of the Hilbert space of a system described by continuous quantum variables. With these codes, continuous variable systems can be used for robust storage and fault-tolerant processing of quantum  information. 

For example, the coded information could reside in the Hilbert space of a single-particle system described by canonical quantum variables $q$ and $p$. In practice, these variables might describe the states of a mode of the electromagnetic field in a high-finesse microcavity, or the state of the center of mass motion of an ion in a trap. Or the continuous Hilbert space could be the state space of a rotor described by an angular variable $\theta$ and its conjugate angular momentum $L$; in practice, these variables might be the phase and charge of a superconducting quantum dot. Our coding scheme can also be applied to a charged particle in a magnetic field. 

Our codes are designed to protect against small errors that occur continually -- diffusive drifts in the values of the canonical variables. The codes are less effective in protecting against large errors that occur rarely. In some settings, we may desire protection against both kinds of errors. One way to achieve that would be to {\em concatenate} our continuous-variable codes with conventional finite-dimensional quantum codes. 

When we consider how to manipulate continuous-variable quantum information fault tolerantly, the issues that arise are rather different than in previous discussions of quantum fault tolerance. With continuous variable codes, propagation of error from one oscillator to another is not necessarily a serious problem. More damaging are processes that amplify a small shift of the canonical variables to a large shift. We have described how to implement a universal set of fault-tolerant quantum gates; with these, harmful error amplification can be avoided as the encoded state is processed. 

Apart from encouraging the intriguing possibility that continuous quantum variables might prove useful for the construction of robust quantum memories and computers, these new quantum codes also have important theoretical applications. In this paper we have discussed an application to the theory of the quantum capacity of the Gaussian quantum channel. Furthermore, quantum codes can be invoked to investigate the efficacy of quantum cryptographic protocols, even in cases where the protocol makes no direct use of the encoded states \cite{shor_jp}. With continuous variable codes, we can demonstrate the security of key distribution protocols based on the transmission of continuous variable quantum information. This application is discussed in a separate paper \cite{qkd_jp}. 

\acknowledgments
We gratefully acknowledge helpful discussions with Isaac Chuang, Sumit Daftuar, David DiVincenzo, Andrew Doherty, Steven van Enk, Jim Harrington, Jeff Kimble, Andrew Landahl, Hideo Mabuchi, Harsh Mathur, Gerard Milburn, Michael Nielsen, and Peter Shor. This work has been supported in part by the Department of Energy under Grant No. DE-FG03-92-ER40701, and by the Caltech MURI Center for Quantum Networks under ARO Grant No. DAAD19-00-1-0374. Some of this work was done at the Aspen Center for Physics.


\begin{references}
\bibitem{shor_9} P.~W. Shor, ``Scheme for reducing decoherence in quantum computer memory,'' Phys. Rev. A {\bf 52}, R2493 (1995).
\bibitem{steane_7} A. Steane, ``Error-correcting codes in quantum theory,'' Phys. Rev. Lett. {\bf 77}, 793 (1996).
\bibitem{lanl_linear} E. Knill, R. Laflamme, and G. Milburn, ``A scheme for efficient quantum computation with linear optics,'' Nature {\bf 409}, 46-52 (2001); ``Efficient linear optics quantum computation,'' quant-ph/0006088.
\bibitem{lanl_thresh} E. Knill, R. Laflamme, and G. Milburn, ``Thresholds for linear optics quantum computation,'' quant-ph/0006120.
\bibitem{knill_higher} E. Knill, ``Non-binary unitary error bases and quantum codes,'' quant-ph/9608048; E. Knill, ``Group representations, error bases and quantum codes,'' quant-ph/9608049.
\bibitem{chau} H.~F. Chau, ``Correcting quantum errors in higher spin systems,'' Phys. Rev. A {\bf 55}, R839 (1997), quant-ph/9610023; H.~F. Chau, ``Five quantum register error correction for higher spin systems,'' Phys. Rev. A {\bf 56}, R1 (1997), quant-ph/9702033.
\bibitem{rains} E.~M. Rains, ``Nonbinary quantum codes,'' quant-ph/9703048. 
\bibitem{higher} D. Gottesman, ``Fault-tolerant quantum computation with higher-dimensional systems,'' Lect. Notes. Comp. Sci. {\bf 1509}, 302 (1999), quant-ph/9802007.
\bibitem{att} A.~R. Calderbank, E.~M. Rains, P.~W. Shor, and N.~J.~A. Sloane, ``Quantum error correction and orthogonal geometry,'' Phys. Rev. Lett. {\bf 78}, 405 (1997), quant-ph/9605005. 
\bibitem{gott_stab} D. Gottesman, ``A class of quantum error-correcting codes saturating the quantum Hamming bound,'' Phys. Rev. A {\bf 54}, 1862 (1996), quant-ph/9604038.
\bibitem{braunstein} S. Braunstein, ``Error correction for continuous quantum variables,'' Phys. Rev. Lett. {\bf 80}, 4084 (1998), quant-ph/9711049. 
\bibitem{lloyd} S. Lloyd and J.~E. Slotine, ``Analog quantum error correction,'' Phys. Rev. Lett. {\bf 80}, 4088 (1998), quant-ph/9711021.
\bibitem{plenio} S. Parker, S. Bose, and M. B. Plenio, ``Entanglement quantification and purification in continuous variable systems,'' Phys. Rev. A {\bf 61}, 32305 (2000), quant-ph/9906098.
\bibitem{zoller} L. M. Duan, G. Giedke, J.~I. Cirac, and P. Zoller, ``Entanglement purification of Gaussian continuous variable quantum states,'' Phys Rev. Lett. {\bf 84}, 4002-4005 (2000), quant-ph/9912017; L. M. Duan, G. Giedke, J.~I. Cirac, and P. Zoller, ``Physical implementation for entanglement purification of Gaussian continuous variable quantum systems,'' Phys. Rev. A {\bf 62}, 032304 (2000), quant-ph/0003116.
\bibitem{cal_shor} A.~R. Calderbank and P.~W. Shor, ``Good quantum error-correcting codes exist,'' Phys. Rev. A {\bf 54}, 1098 (1996), quant-ph/9512032.
\bibitem{steane} A. Steane, ``Multiple particle interference and quantum error correction,'' Proc. Roy. Soc. London, Ser. A {\bf 452}, 2551 (1996), quant-ph/9601029.
\bibitem{ben-or} D. Aharonov and M. Ben-Or, ``Fault-tolerant quantum computation with constant error,'' {\it Proc. 29th Ann. ACM Symp. on Theory of Computing}, p. 176 (ACM, New York, 1998), quant-ph/9611025; D. Aharonov and M. Ben-Or, ``Fault-tolerant quantum computation with constant error rate,'' quant-ph/9906129.
\bibitem{cover} T.~M. Cover and J.~A. Thomas, {\it Elements of Information Theory}, Wiley, New York (1991).
\bibitem{holevo} A.~S. Holevo and R.~F. Werner, ``Evaluating capacities of bosonic Gaussian channels,'' quant-ph/9912067.
\bibitem{lloyd_conj} S. Lloyd, ``The capacity of the noisy quantum channel,'' Phys. Rev. A {\bf 56}, 1613 (1997), quant-ph/9604015.
\bibitem{schumacher} B.~W. Schumacher and M.~A. Nielsen, ``Quantum data processing and error correction,'' Phys. Rev. A {\bf 54}, 2629 (1996), quant-ph/9604022.
\bibitem{barnum} H. Barnum, M.~A. Nielsen, and B. Schumacher, ``Information transmission through a noisy quantum channel,'' Phys. Rev. A {\bf 57}, 4153 (1998), quant-ph/9702049.
\bibitem{ibm} C.~H. Bennett, D.~P. DiVincenzo, J.~A. Smolin, and W.~K. Wootters, ``Mixed state entanglement and quantum error correction,'' Phys. Rev. A {\bf 54}, 3824 (1996), quant-ph/9604024.
\bibitem{shor_smolin} P.~W. Shor and J.~A. Smolin, ``Quantum error-correcting codes need not completely reveal the error syndrome,'' quant-ph/9604006; D.~P. DiVincenzo, P.~W. Shor and J.~A. Smolin, ``Quantum channel capacity of very noisy channels,'' Phys. Rev A {\bf 57}, 830 (1998), quant-ph/9706061.
\bibitem{referee} We thank the anonymous referee for this comment.
\bibitem{reck} M. Reck, A. Zeilinger, H.~J. Bernstein, and P. Bertani, ``Experimental realization of any discrete unitary operator,'' Phys. Rev. Lett. {\bf 73}, 58 (1994). 
\bibitem{shor_ft} P.~W. Shor, ``Fault-tolerant quantum computation,'' {\it Proc. 37th Annual Symp. on Found. of Comp. Sci.}, p. 56 (IEEE, Los Alamitos, CA, 1996), quant-ph/9605011.  
\bibitem{gott_ft} D. Gottesman, ``A theory of fault-tolerant quantum computation,'' Phys. Rev. A {\bf 57}, 127 (1998), quant-ph/9702029.
\bibitem{steane_anc} A. Steane, ``Active stabilization, quantum computation, and quantum state synthesis,'' Phys. Rev. Lett. {\bf 78}, 2252 (1997), quant-ph/9611027.
\bibitem{knill} E Knill, R. Laflamme, W.~H. Zurek, ``Resilient quantum computation: error models and thresholds,'' Proc. Roy. Soc. London, Ser. A {\bf 454}, 365 (1998), quant-ph/9702058.
\bibitem{ike_dan} D. Gottesman and I. Chuang, ``Quantum teleportation is a universal computational primitive,'' Nature {\bf 402}, 390 (1999), quant-ph/9908010.
\bibitem{milburn} S. Schneider, H.~M. Wiseman, W.~J. Munro, and G.~J. Milburn, ``Measurement and state preparation via ion trap quantum computing,'' Fort. der Physik {\bf 46}, 391 (1998), quant-ph/9709042.
\bibitem{mor} P.~O. Boykin, T. Mor, M. Pulver, V. Roychowdhury, and F. Vatan, ``A new universal and fault-tolerant quantum basis,'' Inform. Process Lett. {\bf 75}, 101-107 (2000), quant-ph/9906054.
\bibitem{leung} X. Zhou, D.~W. Leung, and I.~L. Chuang, ``Methodology for quantum logic gate construction,'' Phys. Rev. A {\bf 62}, 052316 (2000), quant-ph/0002039.
\bibitem{kit_purify} A.~Yu. Kitaev, unpublished.
\bibitem{dennis} E. Dennis, Fault-tolerant computation without concatenation, quant-ph/9905027.
\bibitem{nonlinear} V. Giovannetti, S. Mancini, and P. Tombesi, ``Radiation pressure induced Einstein-Podolsky-Rosen paradox,'' quant-ph/0005066, and references therein.
\bibitem{kitaev} A.~Yu. Kitaev, ``Fault-tolerant quantum computation by anyons,'' quant-ph/9707021.
\bibitem{shor_jp} P.~W. Shor and J. Preskill, ``Simple proof of security of the BB84 quantum key distribution protocol,'' Phys. Rev. Lett. {\bf 85}, 441-444 (2000), quant-ph/0003004.
\bibitem{qkd_jp} D. Gottesman and J. Preskill, ``Secure quantum key distribution using squeezed states,'' Phys. Rev. A {\bf 63}, 022309 (2001), quant-ph/0008046. 
\end{references}
\end{document}